\documentclass[prb,twocolumn,amsmath,amssymb,10pt,aps,longbibliography,superscriptaddress,citeautoscript,bibnotes,footinbib]{revtex4-2}

\usepackage[normalem]{ulem} 

\usepackage{feynmp}
\usepackage{amsmath}
\usepackage{graphicx}
\usepackage{multirow}
\usepackage{latexsym}
\usepackage{textcomp}
\usepackage{verbatim}
\usepackage{color}
\usepackage{bm}
\usepackage{subfigure}
\usepackage{everysel}
\usepackage{keyval}
\usepackage{ragged2e}
\usepackage{dsfont}
\usepackage{amssymb}
\usepackage{enumitem}
\usepackage{pstricks}
\usepackage{mathtools}
\usepackage{braket}
\usepackage{slashed}
\usepackage[colorlinks=true]{hyperref}

\usepackage{graphicx}
\usepackage{xcolor}
\usepackage{amsmath}
\usepackage{amsfonts}
\usepackage{bbm}
\usepackage{siunitx}

\usepackage{braket}
\usepackage{slashed}

\newcommand{\bt}[1]{\textbf{\textit{#1}}}
\newcommand{\bp}[1]{\boldsymbol{\Omega}_{\bt{p}}^{#1}}
\newcommand{\pint}{\int \frac{d^3p}{(2 \pi \hbar)^3}}

\begin{document}

\title{Non-divergent Chiral Charge Pumping in Weyl Semimetal}

\author{Min Ju Park}
\thanks{Electronic Address: mandypp@postech.ac.kr}
\affiliation{Department of Physics, Pohang University of Science and Technology, Pohang 37673, Republic of Korea}

\author{Suik Cheon}
\thanks{Electronic Address: enprodigy@postech.ac.kr}
\affiliation{Department of Physics, Pohang University of Science and Technology, Pohang 37673, Republic of Korea}

\author{Hyun-Woo Lee}
\thanks{Electronic Address: hwl@postech.ac.kr}
\affiliation{Department of Physics, Pohang University of Science and Technology, Pohang 37673, Republic of Korea}

\date{\today}

\begin{abstract}
Recent studies suggest that the nonlinear transport properties in Weyl semimetal may be a measurable consequence of its chiral anomaly. Nonlinear responses in transport are estimated to be substantial, because in real materials such as TaAs or Bi$_{1-x}$Sb$_x$, the Fermi level resides near the Weyl nodes where the chiral charge pumping is said to diverge. However, this work presents semiclassical Boltzmann analysis that indicates that the chiral charge pumping is non-divergent even at the zero-temperature limit. We demonstrate that the divergence in common semiclassical calculation scheme is not a problem of the scheme itself, but occurs because a commonly-used approximation of the change in particle number breaks down near the Weyl nodes. Our result suggests the possibility that the nonlinear properties in WSMs can be overestimated, and provides the validity condition for the conventional approximation. We also show the distinct Fermi level dependencies of the chiral magnetic effect and the negative longitudinal magnetoresistance, as a consequence of non-diverging chiral charge pumping.
\end{abstract}

\maketitle
\section{Introduction}

Weyl semimetal (WSM) and its anomalous transport phenomena have been studied intensively~\cite{Xiao1959,Nagaosa2010,Qi2011,Hosur2013,Liang2015,Liang2018,Armitage2018}, as a part of the interest in nontrivial topological properties in topological materials~\cite{Moore2007,Fu2007,Hasan2010}. A WSM has Weyl fermions as its quasiparticle excitations, which are associated with band crossing points (Weyl nodes) that have topologically-protected chiral charges~\cite{Weyl1929,Herring1937,Murakami2007,Burkov2011,Huang2015,Burkov2018,Armitage2018}. However, chiral symmetry in WSM is violated (or chiral current is not conserved) by quantum correction in the presence of non-perpendicular electric field \bt{E} and magnetic field \bt{B}. This violation, referred as Adler-Bell-Jackiw anomaly or chiral anomaly~\cite{Adler1969,Bell1969}, is depicted using a picture of chiral charge pumping (CCP), in which particles are pumped between Weyl nodes of opposite chirality in a direction that is determined by the sign of $\bt{E}\cdot\bt{B}$; the result is an imbalance in the chemical potential between the two Weyl nodes~\cite{Aji2012,Son2013}.

Electron transport properties induced by chiral anomaly have been studied extensively, e.g. chiral magnetic effect (CME)~\cite{Fukushima2008,Kharzeev2009,Son2012,Stephanov2012,Burkov2015prb,Li2016,Cheon2022} and negative longitudinal magnetoresistance (NLMR)~\cite{Nielsen1983,Son2013,Kim2013,Xiong2015,HuangX2015,Cheon2022}, as parts of the methods to experimentally identify the chiral anomaly. However, negative magnetoresistance may arise even in materials that lack Weyl nodes~\cite{Kim2009,Noh2009,Goswami2015,Dos2016,Kikugawa2016,Ong2021}. This recognition naturally leads to a search for other measurable consequences of the chiral anomaly.

Recent theoretical and experimental studies have suggested that WSMs may have nonlinear current, which has nonnegligible or even gigantic magnitude~\cite{Morimoto2016,Shin2017,Nagpal2020,Nandy2021,Li2021,Vashist2021}. For example, a nonlinear current proportional to $\bt{E}^3$ has been reported for time-reversal-symmetry broken (TRSB) WSM~\cite{Shin2017}, and a non-reciprocal current has been reported for inversion-symmetry broken (ISB) WSM~\cite{Morimoto2016}. Such nonlinear transport phenomena can be derived using semiclassical analysis:  the introduction of chiral anomaly to linear transport terms through unbalanced chemical potential brings about higher-order terms in $\bt{E}\cdot\bt{B}$~\cite{Morimoto2016,Shin2017}. In general, the substantial magnitude of such nonlinear responses are closely related to the observation that the amount of the CCP (or the difference of chemical potential) diverges as the Fermi level ($\epsilon_\text{F,0}$) approaches the Weyl nodes. The divergence is often attributed to the diverging Berry curvature at the Weyl nodes. However, some semiclassical calculations~\cite{Kim2014,Morimoto2016,Shin2017} noted that commonly-used calculation schemes may break down close to Weyl nodes.

In this paper, we show that the CCP is non-divergent for both TRSB and ISB WSMs even at the zero-temperature limit of semiclassical analysis. To show this point, we abandon a commonly-used approximation which takes the change of particle density to be proportional to the density of states (DoS). Near the Weyl nodes, where the DoS vanishes, this approximation becomes unreliable. When this approximation is corrected, the CCP becomes non-divergent and is replaced by a weaker singularity. The chemical potential imbalance by the CCP stops being linear to $\bt{E}\cdot\bt{B}$. We have also found the validity condition for the conventional approximation, and shown that the condition is not satisfied in recent experiments with TaAs and  Bi$_{0.96}$Sb$_{0.04}$~\cite{Weng2015,Morimoto2016,Shin2017,Ramshaw2018}.

Lastly, we discuss the distinction between the CME and the NLMR. It is often said that the CME, which is proportional to the magnitude of CCP, cannot be discriminated from the NLMR, which is associated with anisotropic rearrangement of distribution function (or intra-node scattering), because they share similar dependence on \bt{E} and \bt{B}~\cite{Burkov2015prb,Burkov2015iop,Shin2017}. We find that their dependencies on $\epsilon_\text{F,0}$ are distinct when the non-divergent behaviour of the CCP is considered: i.e., the NLMR is divergent because it is a consequence of the Berry curvature effects on equations of motions, but the CME is non-divergent; instead its slope is divergent, because it is directly induced by the CCP. Also, the NLMR and the slope of CME diverge at the point where the Fermi level \textit{after} the CCP is located at the Weyl nodes, which deviates by $\sim40\text{ meV}$ from the intrinsic Weyl node position in TaAs. 

This paper is composed of five sections. Section \ref{sec_model} introduces our model that considers both TRSB and ISB WSMs. Section \ref{sec_ca} gives a brief review of the chiral anomaly from the semiclassical point of view. Section \ref{sec_rnd} presents our main results. We first show our result of non-divergent CME. Then, to investigate the modifications to the nonlinear current terms, we apply the result to other current terms. Section \ref{sec_conclusion} contains concluding remarks.

\section{Model}
\label{sec_model}

\begin{figure}[t] 
\centering{ \includegraphics[width=8.6cm]{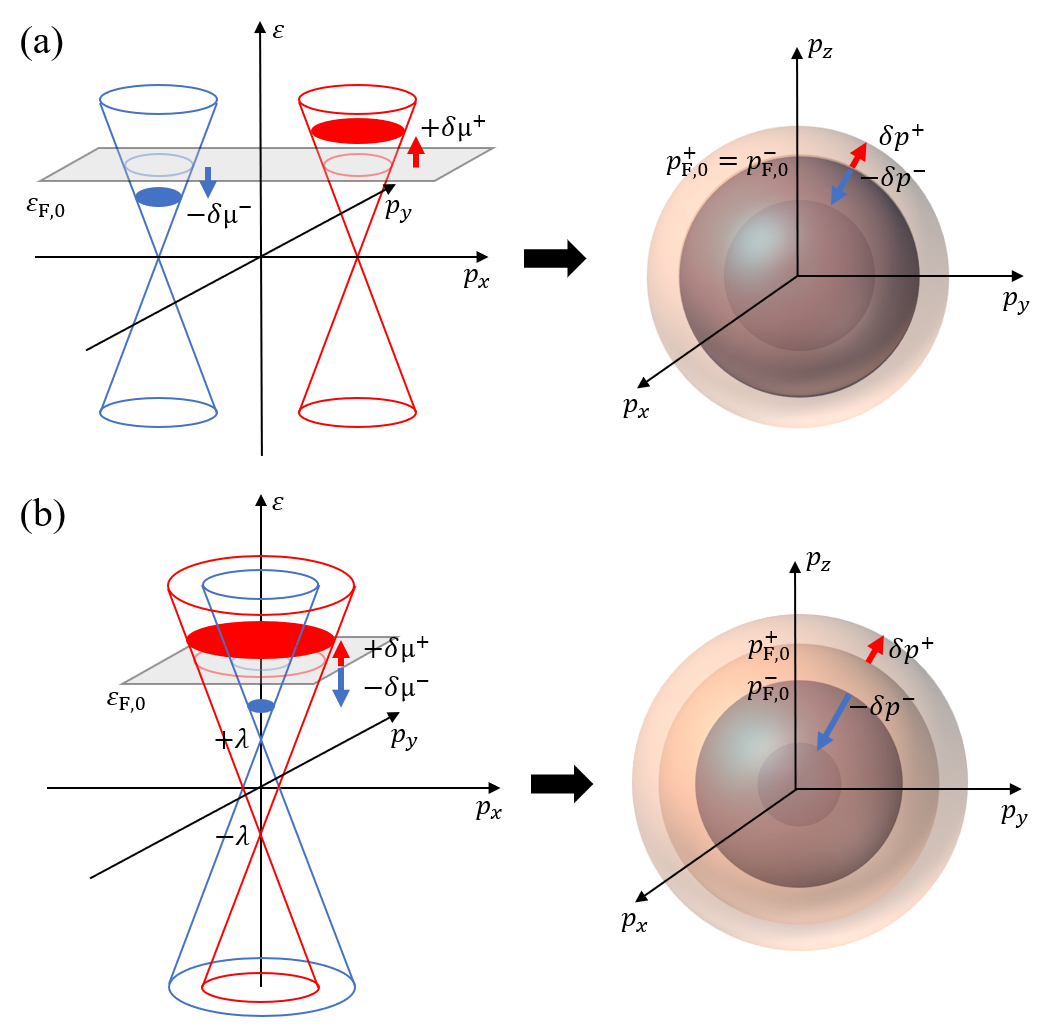}}
\caption{ 
Linear dispersion (left), Fermi surface (right) and their chiral pumping. Red: positive chirality; blue: negative chirality. Fermi surfaces are drawn while locating both Weyl nodes on the same position.
(a) TRSB WSM when $\bt{b}$ is along $p_x$ direction.
(b) ISB WSM. Each node is located at the energy of $\pm \lambda$. Total particle conservation constrains the increased Fermi sphere volume of + chirality to equate with the decreased Fermi sphere volume of - chiralty. As a result, $\left| \delta p^+ \right| < \left|\delta p^-\right|$, so, $\left|\delta \mu^+\right| < \left|\delta \mu^-\right|$ for both (a) and (b) cases.
}\label{Fig1}
\end{figure}

We examine the following low-energy effective model for WSM~\cite{Vazifeh2013,Goswami2013}:
\begin{equation}
H=u\bt{p}\cdot \boldsymbol{\sigma} \otimes \boldsymbol{\tau}_{z}+ \lambda\bt{I} \otimes \boldsymbol{\tau}_{z} + \bt{b}\cdot\boldsymbol{\sigma} \otimes \bt{I},
\label{hamiltonian}
\end{equation}
where $u$ is the Fermi velocity, $\bt{p}$ is the momentum, $\boldsymbol{\tau}_z$ is the chiral index ($+1$ or $-1$), $\boldsymbol{\sigma}$ is a spin index, and $\bt{I}$ is the identity matrix. The first term describes the dispersion near Weyl nodes. The second and the third terms represent different forms of symmetry breaking. The second term represents ISB, and changes the chemical potential of each node by $\lambda$ in opposite directions. The third term represents TRSB, and splits each node in momentum space along $\bt{b}$. Therefore, a TRSB WSM [Fig.~\ref{Fig1}(a)] is accomplished for $\lambda=0$, and a ISB WSM [Fig.~\ref{Fig1}(b)] is accomplished for $\bt{b}=0$. One remark is in order. $H$ in Eq.~\eqref{hamiltonian} contains only two Weyl nodes explicitly, but ISB WSM should contain at least four Weyl nodes: two with chirality $+1$ and two with chirality $-1$. Thus, the model is incomplete. However, it suffices for our transport calculation, in which communication between nodes of the same chirality may be ignored. This problem can be resolved easily by simply multiplying the number of Weyl pairs to responses, which will be calculated in the later sections. In this sense, Eq.~\eqref{hamiltonian} embraces both TRSB and ISB WSM cases.

For the later sections, eigenstate properties of $H$ are calculated here. First, $[\tau_z, H]=0$, and the eigenvalues of $\tau_z$ correspond to negative chiral indices $-\chi=\pm 1$. Therefore, we can reduce the $4 \times 4$ Hamiltonian of Eq.~\eqref{hamiltonian} to the following $2 \times 2$ Hamiltonian of a chirality $\chi$, by replacing $\tau_z$ with the constant $-\chi$,
\begin{equation}
\begin{aligned}
H^{2 \times 2, \chi} = -\chi u\bt{p}\cdot \boldsymbol{\sigma} - \chi \lambda\bt{I} + \bt{b}\cdot\boldsymbol{\sigma} \\
=-\chi u \left( \bt{p} - \chi \frac{\bt{b}}{u} \right) \cdot \boldsymbol{\sigma} - \chi\lambda\bt{I}.
\end{aligned}
\end{equation}
Then for each $\bt{p}$, which is a constant of motion because $[\bt{p},H^{2 \times 2, \chi}]=0$, the above Hamiltonian has two eigenstates with $\boldsymbol{\sigma}$ parallel ($\xi=1$) or anti-parallel ($\xi=-1$) to $-\chi u \left( \bt{p} - \chi \frac{\bt{b}}{u} \right)$. Therefore, two eigenstates can be defined: $\ket{u_{\chi,\xi}(\bt{p})}$ ($\xi=\pm1$) for given $\chi$ and $\bt{p}$. These two eigenstates have energy eigenvalues $\epsilon^{\chi,\xi}(\bt{p})= \xi u \left | \bt{p} - \chi \frac{\bt{b}}{u} \right | -\chi\lambda$. The two resulting bands for given $\chi$ are degenerate at $(\bt{p},\epsilon) = (\chi \frac{\bt{b}}{u}, -\chi\lambda)$, which amounts to a Weyl node. Finally, we obtain the group velocity, $\bt{v}_\bt{p}^{\chi,\xi} = \nabla_{\bt{p}} \epsilon^{\chi,\xi}$, and the Berry field, $\bp{\chi,\xi}= i \hbar^2 \nabla_{\bt{p}} \times \bra{u_{\chi,\xi}(\bt{p})} \nabla_{\bt{p}} \ket{u_{\chi,\xi}(\bt{p})} $, from the dispersion relations and the eigenstates enlightened above;
\begin{equation}
\bt{v}_{\bt{p}}^{\chi,\xi}= \xi u \frac{\bt{p} - \chi \frac{\bt{b}}{u}}{ \left | \bt{p} - \chi \frac{\bt{b}}{u} \right |} , \qquad \bp{\chi,\xi}=\chi \xi \frac{\hbar^2 \left( \bt{p} - \chi \frac{\bt{b}}{u} \right)}{2 \left | \bt{p} - \chi \frac{\bt{b}}{u} \right | ^3} .
\label{vbp}
\end{equation}
For convenience, we shift the $\bt{p}$-coordinate as below to place the Weyl node at the origin,
\begin{equation}
\bt{p} - \chi \frac{\bt{b}}{u} \rightarrow \bt{p}.
\end{equation}  

\section{Chiral Anomaly}
\label{sec_ca}

In quantum field theory, the chiral symmetry of WSM is broken during the ultraviolet regularization process that involves the symmetry-broken integral measure. This violation of the chiral current conservation is expressed as
\begin{equation}
\frac{d N^{\chi}}{d t}  = \chi \frac{e^2}{4\pi^2\hbar^2}\bt{E} \cdot \bt{B},
\label{lpumping}
\end{equation}
where $N^{\chi}$ is the particle number of chirality $\chi$. Equation~\eqref{lpumping} can be obtained alternatively in semiclassical perspective, starting from the Boltzmann equation~\cite{Son2013,Kim2014}. Here, $N^{\chi}$ is defined as
\begin{equation}
N^{\chi}=\sum_{\xi} \int \frac{d^3p}{(2 \pi \hbar)^3}  D^{\chi,\xi}(\bt{p}) f^{\chi, \xi}(\bt{p}),
\label{N}
\end{equation}
where $D^{\chi,\xi}(\bt{p})=1+\frac{e}{\hbar}\bt{B}\cdot\bp{\chi,\xi}$ is a phase-space factor~\cite{Xiao2005}, and $f^{\chi, \xi}(\bt{p})$ indicates the electron occupation, which should be calculated from the Boltzmann equation, at the state with $\bt{p}, \chi,$ and $\xi$. Here, temperature is not considered, because we regard the zero-temperature limit from now on. Since the electron configuration rearrangement within a node does not affect $N^{\chi}$, we take $f^{\chi, \xi}(\bt{p})$ as a step function where electrons are occupied up to the chemical potential. Together with the semiclassical equations of motion for $\dot{\bt{p}}^{\chi,\xi}$ [Eq.~\eqref{eom}], the time-derivative of $N^{\chi}$ in Eq.~\eqref{N} results in Eq.~\eqref{lpumping}, while $\chi$ on the right-hand side of Eq.~\eqref{lpumping} is replaced with $(1/2 \pi \hbar^2) \oint_{S^{\chi,\xi}} \Omega^{\chi,\xi}_\bt{p} \cdot d\bt{S}^{\chi,\xi}=\pm 1$. Here, the integration is done over the Fermi surface $S^{\chi,\xi}$ of chirality $\chi$, and the normal of unit surface $d\bt{S}^{\chi,\xi}$ follows the direction of $\bt{v}_\bt{p}^{\chi,\xi}$ \cite{Sekine2021}. Therefore, the integration results in the value which is independent of $\xi$, and coincides with chiral index $\chi$ defined from the Sec.~\ref{sec_model}.

Now, we present general guide to how Eq.~\eqref{lpumping} leads to chemical potential imbalance. The quantitative calculation will follow in the next part, with a closer look at the approximation used in the process. Equation ~\eqref{lpumping} predicts that $N^{\chi}$ grows/decreases \textit{indefinitely} when $\chi \bt{E} \cdot \bt{B}$ is positive/negative, but in a real solid, scattering between states with $\chi=+1$ and states with $\chi=-1$ prevents the indefinite growth or decrease. When the inter-node scattering rate is characterized by $1/\tau_\text{v}$, the balance between the chiral anomaly and inter-node scattering generates a steady state with
\begin{equation}
\delta N^{+}-\delta N^{-}=\delta N_5= \frac{e^2}{4\pi^2\hbar^2}\bt{E} \cdot \bt{B} \tau_\text{v},
\label{pumping}
\end{equation}
where $\delta N^{\chi}$ is the deviation of $N^{\chi}$ from its equilibrium value with $\bt{E} \cdot \bt{B}=0$. Together with the total charge conservation
\begin{equation}
\delta N^{+}+\delta N^{-}=0,
\label{conservation}
\end{equation}
Eq.~\eqref{pumping} implies that particles are pumped from the negative chiral branch to the positive chiral branch when $\bt{E} \cdot \bt{B} > 0$, and from positive chiral branch to the negative chiral branch when $\bt{E} \cdot \bt{B} < 0$. Particle pumping between opposite chiralities, $\delta N^{+/-}$, determined from Eq.~\eqref{pumping} and Eq.~\eqref{conservation} leads to chemical-potential pumping, which is defined as
\begin{equation}
\mu_5= \frac{\epsilon_\text{F,pump}^+ - \epsilon_\text{F,pump}^-}{2}=\frac{\delta \mu^+ - \delta \mu^-}{2},
\label{mu5}
\end{equation}
where $\epsilon_\text{F,pump}^{\chi}$ is the chemical potential in chiral branch $\chi$ after pumping, and $\delta\mu^{\chi}=\epsilon_\text{F,pump}^{\chi}-\epsilon_\text{F,0}$, where $\epsilon_\text{F,0}$ is the Fermi level before pumping.

\section{Results \& Discussion}
\label{sec_rnd}
\subsection{Non-diverging CME}

CME refers to generation of an electric current induced by the magnetic field, along the field direction~\cite{Fukushima2008}. 
Semiclassical investigation of the CME starts with the following general expression for an electric current in WSM:
\begin{equation}
\bt{j}^\text{tot}=\sum_{\chi}\bt{j}^{\chi}=e \sum_{\chi,\xi} \pint  D^{\chi,\xi}(\bt{p}) f^{\chi,\xi}(\bt{p}) \dot{\bt{r}}_{\bt{p}}^{\chi,\xi}.
\label{current}
\end{equation}
For the CME, we consider chiral pumping effect between opposite nodes only, so the distribution function $f^{\chi,\xi}(\bt{p})$ can be replaced by the equilibrium distribution function $f_0^{\chi,\xi}(\bt{p})=\Theta(\epsilon_\text{F,pump}^{\chi}-\epsilon^{\chi,\xi}(\bt{p}))$, which means neglecting rearrangement of electrons near a node in response to the electromagnetic field. As $\epsilon_\text{F,pump}$ in the step function indicates Fermi level 'after' pumping, $f_0^{\chi,\xi}(\bt{p})$ already contains the effect of the chiral anomaly. The result is the following expression for the CME:
\begin{equation}
\bt{j}^\text{tot}_\text{CME} = \frac{e^2}{4\pi^2\hbar^2} \mu_5 \bt{B}.
\label{CME}
\end{equation}
The magnitude of the CME is determined solely by the value of $\mu_5$. Thus, the subsequent analyses of the CME can be reduced to the calculation of $\mu_5$. 

In many theoretical calculations, $\mu_5$ is said to diverge as $\epsilon_\text{F,0}$ approaches the Weyl node. This inference is drawn from the observation that the DoS vanishes at the Weyl node; specifically, the statement stems from the commonly-used approximation~\cite{Morimoto2016}
\begin{equation}
\delta N^{\chi} \sim (p_\text{F,0}^{\chi})^2 \delta p^{\chi},
\label{deltaNnagaosa}
\end{equation}
where $p_\text{F,0}^{\chi}$ is the Fermi momentum of branch $\chi$ before pumping, and $\delta p^{\chi}$ denotes the change in Fermi momentum. Here, $(p_\text{F,0}^{\chi})^2$ is proportional to the DoS of branch $\chi$. Therefore, Eq.~\eqref{deltaNnagaosa} implies that finite $\delta N^{\chi}$ induces infinite $\delta p^{\chi}$ when $\epsilon_\text{F,0}$ is at the Weyl node ($p_\text{F,0}^{\chi}=0$). Then $\mu_5$ diverges due to the linear relation between energy and momentum. Also, because of this linear relation together with Eq.~\eqref{pumping}, the resultant $\mu_5$ from Eq.~\eqref{deltaNnagaosa} becomes linear to $\bt{E} \cdot \bt{B}$, which is a general form of chiral pumping in WSM~\cite{Shin2017,Li2016}.

However, the above analysis is seriously modified when the approximation in Eq.~\eqref{deltaNnagaosa} is replaced by the exact expression
\begin{equation}
\delta N^{\chi} = \frac{1}{2 \pi^2 \hbar^3} \int_{p_\text{F,0} ^{\chi}}^{p_\text{F,pump}^{\chi}} p^2 dp.
\label{deltaNfull}
\end{equation}

\begin{figure}[t] 
\centering{ \includegraphics[width=8.6 cm]{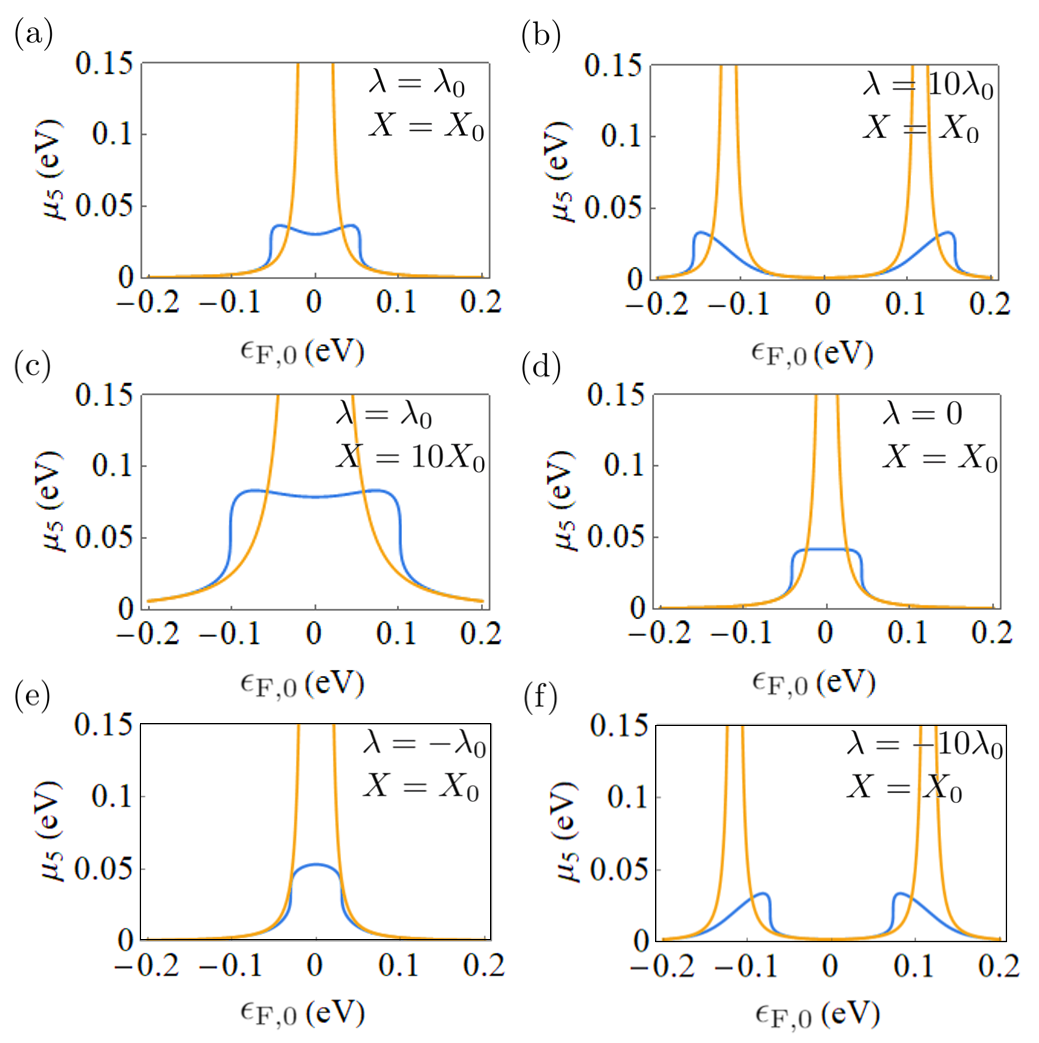} }
\caption{ 
$\mu_5$ behaviours to $\epsilon_\text{F,0}$ from the exact $\delta N^{\chi}$ [Eq.~\eqref{deltaNfull}] (blue lines), and from the conventional approximation of $\delta N^{\chi}$ [Eq.~\eqref{deltaNnagaosa}] (yellow lines). $\lambda_{0} = 11.5 \text{ meV}$~\cite{Weng2015} and $X_{0}=(41.9 \text{ meV})^3$~\cite{Ramshaw2018,Morimoto2016} are the parameters for TaAs, which are obtained from a recent experiment and frist-principle studies of TaAs. $\lambda=0$ represents the case of TRSB WSM; the rest represents the case of ISB WSM. $\lambda>0$ represents the case where + chirality node is located at lower energy; $\lambda<0$ represents the opposite case. 
}\label{Fig2}
\end{figure}

Equation~\eqref{deltaNfull} identifies that $\delta N^{\chi}$ is the volume change of the Fermi sphere (Fig.~\ref{Fig1}, right), and demonstrates that $\delta p^{\chi}=p_\text{F,pump}^{\chi}-p_\text{F,0} ^{\chi}$ does not diverge even when $p_\text{F,0}^{\chi}=0$, because diverging $\delta p^{\chi}$ implies diverging $\delta N^{\chi}$ according to Eq.~\eqref{deltaNfull}. Together with Eqs.~\eqref{pumping} and \eqref{conservation}, Eq.~\eqref{deltaNfull} determines $p_\text{F,pump}^{\chi}$ in the presence of the chiral charge pumping and one obtains the consequent chemical potential after the chiral charge pumping,
\begin{equation}
\epsilon_\text{F,pump}^{\pm}= \sqrt[3]{(\epsilon_\text{F,0} \pm \lambda)^3 \pm X} \mp \lambda,
\label{epsilonpm}
\end{equation}
where $X=\frac{3}{4} u^3 \tau_\text{v} \hbar e^2 (\bt{E} \cdot \bt{B})$. Now, $\mu_5$ can be obtained from Eq.~\eqref{epsilonpm} as in Eq.~\eqref{mu5}. The resulting true $\mu_5$ (Fig.~\ref{Fig2}, blue lines) as a function of $\epsilon_\text{F,0}$ differs greatly from $\mu_5$ obtained from the approximation [Eq.~\eqref{deltaNnagaosa}] (Fig.~\ref{Fig2}, yellow lines). 

The behaviour of $\mu_5$ to $\epsilon_\text{F,0}$ depends on the values of $\lambda$ and $X$ (Fig.~\ref{Fig2}). The exact expression for $\delta N^{\chi}$ predicts plateau-shaped plots (blue lines), while the approximation predicts diverging plots (yellow lines). The effects of parameters $\lambda$ and $X$ on the slope of $\mu_5$ are examined by varying $\lambda$ and $X$. When $\lambda$ increases by factor 10 ($\lambda=10\lambda_0$, $X=X_0$) so that $\lambda>X^{1/3}$ [Fig.~\ref{Fig2}(b)], $\mu_5$ between the two Weyl nodes is suppressed so that the blue and yellow lines agree with each other. Near each Weyl node, however, the difference between the blue and yellow lines persists since the blue line forms highly asymmetric-shaped profile. When $X$ increases by factor 10 ($\lambda=\lambda_0$, $X=10X_0$) so that $\lambda<X^{1/3}$ [Fig.~\ref{Fig2}(c)], $\mu_5$ between the two Weyl nodes gets flattened more. In the extreme case of $\lambda=0$ [Fig.~\ref{Fig2}(d)], which amounts to the TRSB WSM, the plateau shape gets entirely flattened at the height of $\sqrt[3]{X}$ with the width of $2\sqrt[3]{X}$.

Formation of this plateau is closely linked to the realization of the infinitely large slope of $\mu_5$ before $\epsilon_\text{F,0}$ reaches the Weyl nodes. To understand why $\partial \mu_5 / \partial \epsilon_\text{F,0}$ diverges, we obtain from Eq.~\eqref{epsilonpm},
\begin{equation}
\frac{\partial \epsilon_\text{F,pump}^{\pm}}{\partial \epsilon_\text{F,0}}= \frac{(\epsilon_\text{F,0} \pm \lambda)^2}{((\epsilon_\text{F,0} \pm \lambda)^3 \pm X)^{\frac{2}{3}}},
\end{equation}
which approaches infinity at $\epsilon_\text{F,0}=\mp(\lambda+\sqrt[3]{X})$. For these $\epsilon_\text{F,0}$, either chemical potentials after CCP ($\epsilon_\text{F,pump}^{+}$ or $\epsilon_\text{F,pump}^{-}$) is located at the Weyl node. If $\lambda \geq 0$ [Fig.~\ref{Fig2}(a - d)], the width of the plateau part are wider than the width between two Weyl nodes $2\lambda$ by $2\sqrt[3]{X}$. The magnitude of CME is expected to be stable in this $\epsilon_\text{F,0}$ range (about $0.1\text{ eV}$ wide even though $\lambda=0$). This widening of plateau occurs because the + chirality node exists at lower energy level than the - chirality node. These two nodes exchange their relative energies with each other when $\lambda$ becomes negative [Fig.~\ref{Fig2}(e, f)], then the infinite slope of $\mu_5$ is realized in between the two Weyl nodes. At this time, the width of plateau can vanish if $\lambda+\sqrt[3]{X}=0$, unlike positive $\lambda$ cases. Also, when $\lambda+\sqrt[3]{X}<0$ [Fig.~\ref{Fig2}(f)], the positive-infinity slope appears near the higher energy Weyl node and the negative-infinity slope near the lower energy node.

Returning to the overall shape of Fig.~\ref{Fig2}, the deviation between the blue and the yellow lines is clear near the Weyl nodes, but negligible when $\epsilon_\text{F,0}$ is sufficiently far from them. We searched for the condition where the approximation [Eq.~\eqref{deltaNnagaosa}] is valid. We expand Eq.~\eqref{epsilonpm} in a series of $X$ in the case $\left | \epsilon_\text{F,0} \pm \lambda \right | \gg \sqrt[3]{X}$ to obtain
\begin{equation}
\begin{aligned}
\epsilon_\text{F,pump}^{+} &=\epsilon_\text{F,0} + \frac{X}{3 (\epsilon_\text{F,0} + \lambda)^2} - \frac{X^2}{9 (\epsilon_\text{F,0} + \lambda)^5} + \cdots \\
\epsilon_\text{F,pump}^{-} &=\epsilon_\text{F,0} - \frac{X}{3 (\epsilon_\text{F,0} - \lambda)^2} - \frac{X^2}{9 (\epsilon_\text{F,0} - \lambda)^5}  - \cdots.
\end{aligned}
\label{expansion}
\end{equation}
Taking $\epsilon_\text{F,pump}^{\pm}$ in Eq.~\eqref{expansion} to the first order of $X$ (or $\bt{E} \cdot \bt{B}$), then evaluating $\mu_5$ renders exactly the same form of $\mu_5$ as obtained from the approximation [Eq.~\eqref{deltaNnagaosa}]. Therefore, to justify the conventional approximation, the following condition should be satisfied,
\begin{equation}
\left | \epsilon_\text{F,0} \pm \lambda \right | \gg \sqrt[3]{X}.
\label{condition}
\end{equation}

Now that the condition is found, the next step is to check whether it is satisfied in real materials and real experimental conditions. For this, we estimate the parameters $\epsilon_\text{F,0}$, $\lambda$, and $X=\frac{3}{4} u^3 \tau_\text{v} \hbar e^2 (\bt{E} \cdot \bt{B})$ for TaAs and Bi$_{0.96}$Sb$_{0.04}$, which are ISB and TRSB WSM, respectively. $\epsilon_\text{F,0}$ is usually determined from the calculation of the band structure. For TaAs, $\epsilon_\text{F,0}=9.5 \text{ meV}$, $u=4 \times 10^5 \text{ m/s}$, and $\lambda = 11.5 \text{ meV}$~\cite{Weng2015}. For Bi$_{0.96}$Sb$_{0.04}$, $\epsilon_\text{F,0}=10 \text{ meV}$, and $u=10^5 \text{ m/s}$\cite{Shin2017}. Then $\tau_\text{v}$ is roughly estimated from the empirical Drude conductivity $\sigma_\text{D}$, with $\bt{B}$ set to $5 \text{ T}$ and $\bt{E}$ set to $10^3 \text{ V/m}$, with the same direction. Firstly, when $\sigma_\text{D} \sim 2 \times 10^6 \text{ }\Omega^{-1} \text{m}^{-1}$ \cite{Ramshaw2018}, and $\tau_\text{v} / \tau_\text{a}$ is estimated as in~\cite{Morimoto2016}, TaAs gives $\sqrt[3]{X} = 41.9 \text{ meV}$, which is larger than $\left |\epsilon_\text{F,0} \pm \lambda \right |= [2.0 \text{ meV}, 21.0 \text{ meV}]$. Secondly, when $\sigma_\text{D} \sim 3.3 \times 10^5 \text{ }\Omega^{-1} \text{m}^{-1}$, and $\tau_\text{v} / \tau_\text{a}$ is estimated as in \cite{Shin2017}, Bi$_{0.96}$Sb$_{0.04}$ gives $\sqrt[3]{X} = 22.3 \text{ meV}$. Here as well $\sqrt[3]{X}$ is larger than $\left |\epsilon_\text{F,0} \pm \lambda \right |=10 \text{ meV}$. These materials are chosen to be a representative experimental platform of WSM, because of the substantial magnitude of chiral anomaly ($\epsilon_\text{F,0}$ near the Weyl node). However, the condition for the substantial magnitude of chiral anomaly opposes Eq.~\eqref{condition}, which is the condition to legitimate the conventional approach of considering $\mu_5$ to the first order of $\bt{E} \cdot \bt{B}$.  We remark that the above parameters are in the region where the Boltzmann formalism remains valid, which is $\left | \epsilon_\text{F,0} \pm \lambda \right | \gg \frac{\hbar}{\tau_{v}}$.

\subsection{Other Current Terms}

Consideration of the chiral anomaly also affects other transport responses, which come from a nonequilibrium distribution function near a node~\cite{Takasan2021}. The chiral anomaly renders higher-order terms in $\bt{E} \cdot \bt{B}$ to these nonequilibrium responses as a result of the $\bt{E} \cdot \bt{B}$ dependency in $\epsilon_\text{F,pump}^{\pm}$~\cite{Morimoto2016,Shin2017}, so non-diverging chiral pumping demonstrated above is expected to modify the nonlinear currents of WSMs.

We start from the semiclassical approach to the current [Eq.~\eqref{current}]. Here, $f^{\chi,\xi}(\bt{p})$ is modified as below, with the intra-node scattering time $\tau_\text{a}$.
\begin{equation}
f^{\chi,\xi}(\bt{p}) \simeq f_0^{\chi,\xi}(\bt{p}) - \tau_\text{a}(\dot{\bt{p}}^{\chi,\xi}\cdot \nabla_{\bt{p}})f_0^{\chi,\xi}(\bt{p}).
\label{rtapprox}
\end{equation}
The previous study limited $f^{\chi,\xi}(\bt{p})$ in Eq.~\eqref{current} to $f_0^{\chi,\xi}(\bt{p})$, but here we consider anisotropic rearrangement of electrons within a node of $\chi$ and $\xi$, caused by electromagnetic fields, to the first-order scattering. The effect of chiral pumping is included in $f_0^{\chi,\xi}(\bt{p})$, combined with $\sum_{\chi}$ in Eq.~\eqref{current}, by using $\epsilon_\text{F,pump}^{\chi}$ [Eq.~\eqref{epsilonpm}] instead of $\epsilon_\text{F,0}$. In addition, $\dot{\bt{r}}$ and $\dot{\bt{p}}$ from Eq.~\eqref{current} and Eq.~\eqref{rtapprox} are replaced by the semiclassical equations of motion with Berry curvature effects~\cite{Sundaram1999,Son2013}:
\begin{equation}
\begin{aligned}
\dot{\bt{r}}_{\bt{p}}^{\chi,\xi}&= D^{\chi,\xi}(\bt{p})^{-1}[\bt{v}_\bt{p}^{\chi,\xi} + \frac{e}{\hbar} \bt{E} \times \bp{\chi,\xi} + \frac{e}{\hbar} (\bp{\chi,\xi}\cdot \bt{v}_\bt{p}^{\chi,\xi})\bt{B}],  \\
\dot{\bt{p}}^{\chi,\xi}&= D^{\chi,\xi}(\bt{p})^{-1}[e\bt{E} + e \bt{v}_\bt{p}^{\chi,\xi} \times \bt{B} + \frac{e^2}{\hbar} (\bt{E}\cdot \bt{B})\bp{\chi,\xi}].
\end{aligned}
\label{eom}
\end{equation}
This way, the semiclassical approach separates the step of considering the Berry curvature effect [Eq.~\eqref{eom}] from the step of considering the chiral pumping effect, and this separation simplifies tracking of the cause of each response term. Finally, we consider a magnetic field sufficiently smaller than the characteristic magnetic field above which the Landau level quantization becomes relevant, so the result is expressed up to the second order of $\bt{B}$. The consequent current terms are
\begin{equation}
\bt{j}^\text{tot}=\bt{j}^\text{tot}_\text{CME}+\bt{j}^\text{tot}_\text{D}+\bt{j}^\text{tot}_\text{MR}+\bt{j}^\text{tot}_\text{Hall},
\label{cur1}
\end{equation}
where $\bt{j}^\text{tot}_\text{CME}$ is given in Eq.~\eqref{CME}, and
\begin{equation}
\begin{aligned}
\bt{j}^\text{tot}_\text{D}&=\sigma_\text{D} \bt{E},\\
\bt{j}^\text{tot}_\text{MR}&=\frac{\sigma_\text{MR}}{\text{B}_\text{0}^2} \left( \frac{7}{8}(\bt{E} \cdot \bt{B}) \bt{B} + \frac{1}{8} B^2\bt{E} \right),\\
\bt{j}^\text{tot}_\text{Hall}&=\frac{\sigma_\text{Hall}}{\text{E}_\text{0}\text{B}_\text{0}^2}(\bt{E} \cdot \bt{B}) \bt{B} \times \bt{E}.
\end{aligned}
\label{cur2}
\end{equation}
If we set $\text{B}_\text{0}=1\text{ T}$ and $\text{E}_\text{0}=10^3 \text{ V/m}$, then
\begin{equation}
\begin{aligned}
\sigma_\text{D}&= \frac{1}{6}e^2 \frac{u\tau_\text{a}}{\pi^2 \hbar^3} \sum_{\chi}(p_\text{F,pump}^{\chi})^2 \\
\frac{\sigma_\text{MR}}{[\text{T}^2]} &=\frac{1}{15} e^4 \frac{u\tau_\text{a}}{\pi^2 \hbar}  \sum_{\chi} \frac{1}{(p_\text{F,pump}^{\chi})^2}\\
\frac{\sigma_\text{Hall}}{[10^3 \text{ V/m} \cdot \text{T}^2]}&= \frac{1}{80} e^5 \frac{\tau_\text{a}}{\pi^2} \sum_{\chi} \frac{\chi}{(p_\text{F,pump}^{\chi})^4},
\end{aligned}
\label{currents}
\end{equation}
where $(p_\text{F,pump}^{\pm})^2=(\epsilon_\text{F,pump}^{\pm} \pm \lambda)^2/u^2$. Details about how each term of equations of motion leads to each current terms are given in Appendix.~\ref{sec_appendix}. The quantities in Eq.~\eqref{currents} depend on $X$ ($\sim \bt{E}\cdot\bt{B}$): i.e., they include nonlinear responses. Also, only $\bt{j}^\text{tot}_\text{CME}$ comes from the first term $f_0^{\chi,\xi}(\bt{p})$ in Eq.~\eqref{rtapprox} that is of the zeroth-order in $\tau_\text{a}$, and the others come from the second term $-\tau_\text{a}(\dot{\bt{p}}^{\chi,\xi}\cdot \nabla_{\bt{p}})f_0^{\chi,\xi}(\bt{p})$ in Eq.~\eqref{rtapprox} that is of the first-order in $\tau_\text{a}$.
\begin{figure}[t] 
\centering{ \includegraphics[width=8.6 cm]{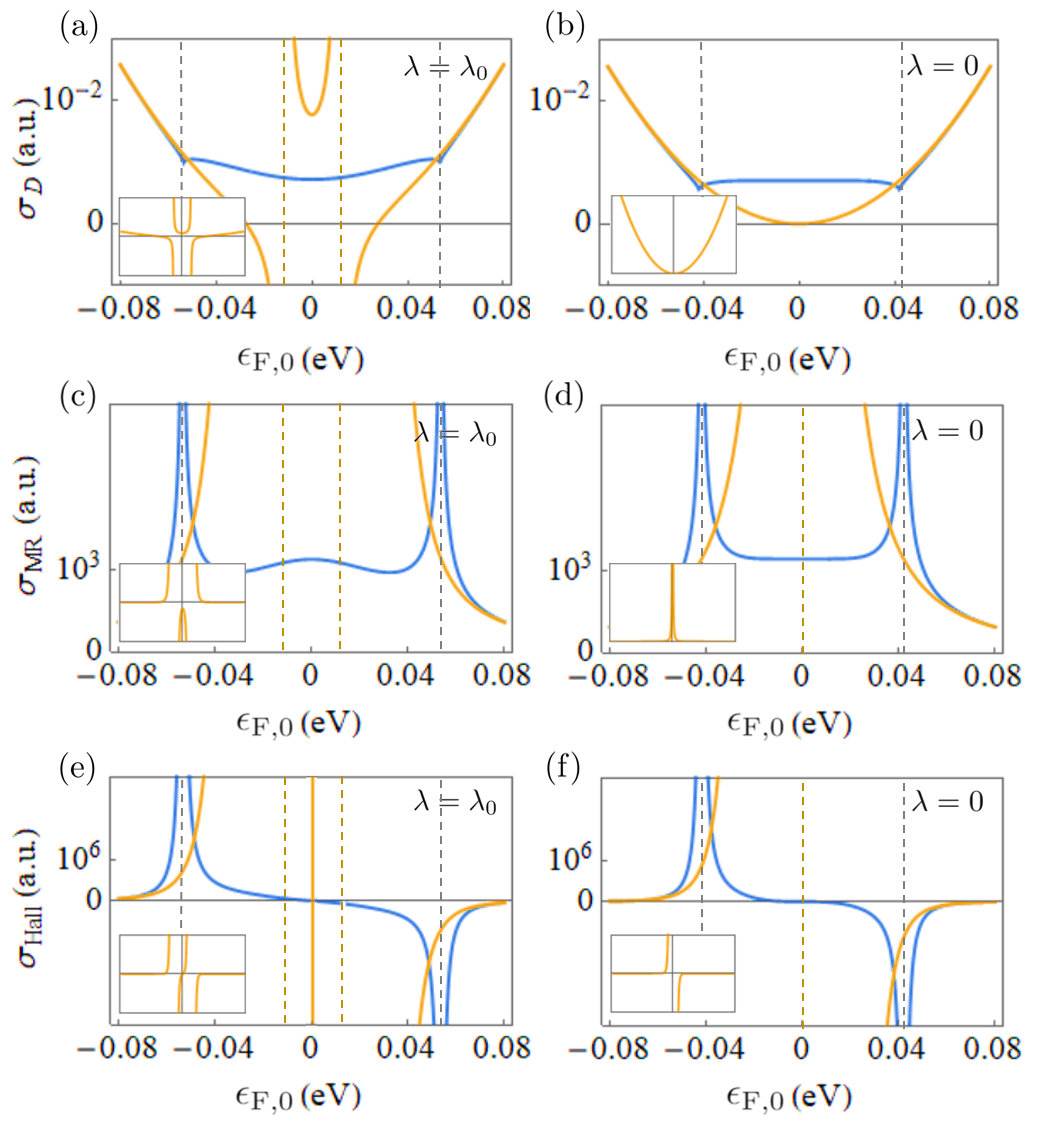} }
\caption{ 
$\sigma_\text{D}$, $\sigma_\text{MR}$, and $\sigma_\text{Hall}$ in their arbitrary unit, where $\frac{1}{6}e^2 \frac{\tau_\text{a}}{u\pi^2 \hbar^3}=1$, $\frac{1}{15} e^4 \frac{u^3\tau_\text{a}}{\pi^2 \hbar}[\text{T}^2]=1$, and $\frac{1}{80} e^5 \frac{u^4\tau_\text{a}}{\pi^2}[10^3 \text{ V/m} \cdot \text{T}^2]=1$, respectively. Each quantity includes nonlinear responses. The results from the precise expression of $\delta N^{\chi}$ [Eq.~\eqref{deltaNfull}] (blue lines), and from the conventional approximation of $\delta N^{\chi}$ [Eq.~\eqref{deltaNnagaosa}] (yellow lines) are compared for both ISB ($\lambda=\lambda_\text{0}$, $X=X_\text{0}$) and TRSB ($\lambda=0$, $X=X_\text{0}$) cases. (a), (c), (e): ISB case; (b), (d), (f): TRSB case.  The yellow lines diverge at $\epsilon_\text{F,0}=\pm\lambda$, which is the Weyl node level \textit{before} chiral pumping, whereas the blue lines diverge or make kink at $\epsilon_\text{F,0}=\pm(\lambda+\sqrt[3]{X})$, where $\epsilon_\text{F,pump}^{\pm}$ is located at one of the Weyl nodes. Insets: the overall shape of yellow lines on expanded y-scale and the same x-scale. The brown and the gray dashed-line are placed to diverging or kinked points of the yellow and the blue lines, respectively.The brown (inner) dashed line is located at $\pm\lambda$, and the gray (outer) dashed line at $\pm(\lambda+\sqrt[3]{X})$.
}\label{Fig3}
\end{figure}

$\sigma_\text{D}$, $\sigma_\text{MR}$, and $\sigma_\text{Hall}$ depends on $\epsilon_\text{F,0}$ (Fig.~\ref{Fig3}). The blue lines in Fig.~\ref{Fig3} are obtained by evaluating $\epsilon_\text{F,pump}^{\pm}$ by using Eq.~\eqref{deltaNfull}, whereas the yellow lines are obtained by evaluating $\epsilon_\text{F,pump}^{\pm}$ by using the conventional approximation, Eq.~\eqref{deltaNnagaosa}. The yellow lines diverge at $\epsilon_\text{F,0}=\pm\lambda$, which is the Weyl node level \textit{before} chiral pumping, while the blue lines diverge or make kink at $\epsilon_\text{F,0}=\pm(\lambda+\sqrt[3]{X})$, where $\epsilon_\text{F,pump}^{\pm}$ is located at one of the Weyl nodes. 

Similar to Fig.~\ref{Fig2}, the blue lines get nearly flat between the Weyl nodes, and become entirely flat for TRSB cases. To explain the shape of the divergence near the nodes, we should keep in mind that the yellow lines are nothing but the expansion of the blue lines to the zeroth and first orders of $X/(\epsilon_\text{F,0} \pm \lambda)^3$ [Eq.~\eqref{expansion}], where $X \sim \bt{E}\cdot\bt{B}$. In this series expansion, terms with one higher order of $X$ have three lower orders of $(\epsilon_\text{F,0}\pm\lambda)$. Thus, each even and odd order term of the X series shows alternating symmetric and anti-symmetric divergence. From the previous part, we checked that the condition Eq.~\eqref{condition} does not hold for TaAs, so the first order term in $X$ determines that the divergence of the yellow lines is anti-symmetric. In contrast, the blue lines include all the higher powers making the divergence near the node symmetric. For TRSB cases, all non-reciprocal response terms (odd powers of $X$ in $\sigma_\text{D}$ and $\sigma_\text{MR}$, and even powers of $X$ in $\sigma_\text{Hall}$) are eliminated, so the divergences of $\sigma_\text{D}$ and $\sigma_\text{MR}$ become symmetric.

Considering the blue lines in Fig.~\ref{Fig3}, the biggest difference between $\sigma_\text{D}$ and the others ($\sigma_\text{MR}$ and $\sigma_\text{Hall}$) is that the former has a kink while the latter have divergences. The difference comes from whether or not each current term is derived directly from the Berry curvature effects on the equations of motion (e.g., anomalous velocity). To elaborate, the term $\bt{j}_\text{D}^\text{tot}$ comes from $\bt{v}_\bt{p}^{\chi,\xi}$ of $\dot{\bt{r}}_\bt{p}^{\chi,\xi}$ and $e\bt{E}$ of $\dot{\bt{p}}^{\chi,\xi}$, which are not the Berry curvature effects from Eq.~\eqref{eom}. Hence, the nonlinearity of $\sigma_\text{D}$ stems solely from the chiral pumping, which this work shows is not divergent. Different from $\bt{j}_\text{D}^\text{tot}$, $\bt{j}_\text{MR}$ and $\bt{j}_\text{Hall}$ come from the terms that contain Berry curvature from Eq.~\eqref{eom}. Berry curvature is diverging at the Weyl nodes [Eq.~\eqref{vbp}], so their divergences cannot be eliminated by non-diverging chiral pumping. For completeness, we remark that the semiclassical formulation on which our analysis is based may require corrections that go beyond the semiclassical formulation when Fermi level lies at the Weyl node and the Fermi surface area shrinks to zero~\cite{Kim2014}.

Some studies have attributed the nonlinear response to the CCP, only counting the lowest nonlinear current term with the approximation [Eq.~\eqref{deltaNnagaosa}]. For example, one may regard the first order of $X$ ($\sim \bt{E} \cdot \bt{B}$) from $\sigma_\text{D}$ as a major contribution from nonreciprocal response of ISB WSM~\cite{Morimoto2016}, or one may incorporate the second order of $X$ from $\sigma_\text{D}$ in the nonlinear response of TRSB WSM~\cite{Shin2017}. However, such consideration is legitimate only in a system in which Eq.~\eqref{condition} holds, because $\sigma_\text{D}$ has further higher-order terms:
\begin{equation}
\begin{aligned}
\sigma_\text{D}= \frac{1}{6}e^2 \frac{\tau_\text{a}}{u \pi^2 \hbar^3} [  2(\epsilon_\text{F,0}^2+\lambda^2)+\frac{4}{3}\frac{\lambda}{(\lambda^2-\epsilon_\text{F,0}^2)}X \\
-\frac{1}{9}\left ( \frac{1}{(\lambda+\epsilon_\text{F,0})^4}+\frac{1}{(\lambda-\epsilon_\text{F,0})^4} \right )X^2 \\
+\frac{4}{81}\left ( \frac{1}{(\lambda+\epsilon_\text{F,0})^7}+\frac{1}{(\lambda-\epsilon_\text{F,0})^7} \right )X^3 + \cdots ].
\end{aligned}
\label{drudeseries}
\end{equation}
Here, the non-reciprocal responses (odd-power of X) change their signs when the Fermi level crosses the Weyl node. Consequently, although each higher-order term is diverging, divergences of $\sigma_\text{D}$ at $\epsilon_\text{F,0}=\pm \lambda$ are suppressed as a whole, so kink shapes arise; i.e., nonlinear responses calculated from a first few terms can be overestimated when $\epsilon_\text{F,0}$ is located close to one of the Weyl nodes, even though such WSMs have been chosen to be good experimental systems, as previously mentioned. Hence, for such materials, total particle conservation [Eq.~\eqref{deltaNfull}] must be considered precisely.

\section{Concluding Remarks}
\label{sec_conclusion}

Starting from the model of WSM in which either inversion symmetry or time reversal symmetry is broken, we have calculated electron transport properties using the semiclassical Boltzmann theory. Our work has pointed out that CME (or CCP) does not diverge around the Weyl nodes even in case level broadening effect from impurity scattering is not considered or at zero temperature. This deviation from the conventional semiclassical calculation that predicts divergence occurs because we treat the particle number change near the Weyl nodes accurately. Our modified plot of CME with respect to $\epsilon_\text{F,0}$ is plateau-shaped, which cannot be expected from the other factors suppressing divergence, such as finite temperature or impurity scattering. Here, $\partial \mu_5/\partial \epsilon_\text{F,0}$, rather than $\mu_5$, goes infinity at the point distance $\sqrt[3]{X}$ from the Weyl node, which corresponds to the Weyl node \textit{after} CCP. Among few experimental data of CME with respect to $\epsilon_\text{F,0}$, Ref.~\cite{Zhang2016} showed that the CME coefficient diverges near a Weyl node. We point out that these data cannot be used to support our work, because $\bt{E}\cdot\bt{B}$ is not big enough to seriously violate the condition for the conventional approximation. Also, Ref.~\cite{Zhang2016} extracted a CME coefficient without considering the chemical potential dependence of dissipative currents which should diverge as Weyl nodes are approached. 

The chiral anomaly effect is regarded to be prominent when $\epsilon_\text{F,0}$ is close to the Weyl nodes, so materials with such property (e.g., TaAs, Bi$_{0.96}$Sb$_{0.04}$) are commonly used in experiments. However, we found that this condition - $\epsilon_\text{F,0}$ near the Weyl nodes - contradicts the validity condition for the conventional approximation. In such materials, nonlinear current terms can be overestimated, so we suggest use of the particle-number change in its intact form. This way, the behaviours of current terms in their full series to electromagnetic fields are modified: the singularities of Drude conductivity weakens to form kinks, whereas other terms (MR and Hall term in this paper) diverge; i.e., CME and NLMR show distinct shapes of plots with respect to $\epsilon_\text{F,0}$. The differences in response occur because the nonlinear terms from Drude conductivity originate solely from CCP, which we found to be non-divergent, whereas other terms stem from Berry curvature effects on electrons' motions even before the CCP is considered. Also, we note that the kink shape in the Drude term and the divergence in other terms appear not when $\epsilon_\text{F,0}$ is located at the Weyl nodes, but when $\epsilon_\text{F,pump}$ is located at the Weyl nodes.

The authors acknowledges support from National Research Foundation of Korea (NRF) grant funded by the Korea government (MSIT) (no. 2020R1A2C2013484). We thank J. M. Lee for valuable conversations.
 
\appendix
\section{Current Calculations}
\label{sec_appendix}
This part shows how Eq.~\eqref{current} leads to Eq.~\eqref{cur1}-\eqref{currents}. The analytic process in case $\bt{B} \parallel \bt{E}$ can be found in detail from \cite{Shin2017}. This part shows which terms in equations of motion result in each current terms in Eq.~\eqref{cur1}.

As stated from 'Chiral Magnetic Effect' part of Section \ref{sec_rnd}, the first term from Eq.~\eqref{rtapprox} gives Eq.~\eqref{CME}. Replacing $f^{\chi,\xi}(\bt{p})$ with the second term from Eq.~\eqref{rtapprox}, then putting Eq.~\eqref{eom} to Eq.~\eqref{current} gives following equation of first-order scattering current,
\begin{equation}
\begin{aligned}
    \bt{j}^{\chi,1}= & e\tau_a \pint  \left [  \bt{v}_\bt{p}^{\chi,\xi} + \frac{e}{\hbar} \bt{E} \times \bp{\chi,\xi} + \frac{e}{\hbar} (\bp{\chi,\xi}\cdot \bt{v}_\bt{p}^{\chi,\xi})\bt{B} \right ] \\
    & D^{\chi,\xi-1} \left [ e\bt{E} + e \bt{v}_\bt{p}^{\chi,\xi} \times \bt{B} + \frac{e^2}{\hbar} (\bt{E}\cdot \bt{B})\bp{\chi,\xi}\right ] \cdot \nabla_{\bt{p}} f_0^{\chi,\xi},
\end{aligned}
\end{equation}
where the $\bt{p}$ dependencies of $D^{\chi,\xi}$ and $f_{0}^{\chi,\xi}$ are omitted for brevity. For the calculation, $D^{\chi,\xi}(\bt{p})^{-1}$ is series expanded, the relation $\nabla_{\bt{p}} f_{0}^{\chi,\xi} =  \bt{v}_{\bt{p}}^{\chi,\xi}\frac{\partial f_{0}^{\chi,\xi}}{\partial \epsilon^{\chi,\xi}} $ is used, and the result is expressed up to the second order of $\bt{B}$. The overall results are given from Eq.~\eqref{cur1}-\eqref{currents}, except that $\bt{j}^{\chi}$, instead of $\bt{j}^\text{tot}$ is obtained here before we perform summation over chirality to get total currents.
\begin{equation}
\bt{j}^{\chi,1}=\bt{j}^{\chi}_\text{D}+\bt{j}^{\chi}_\text{MR}+\bt{j}^{\chi}_\text{Hall},
\label{Acur1}
\end{equation}
where
\begin{equation}
\begin{aligned}
\bt{j}^{\chi}_\text{D}&=\sigma_\text{D}^{\chi} \bt{E},\\
\bt{j}^{\chi}_\text{MR}&=\frac{\sigma_\text{MR}^{\chi}}{\text{B}_\text{0}^2} \left( \frac{7}{8}(\bt{E} \cdot \bt{B}) \bt{B} + \frac{1}{8} B^2\bt{E} \right),\\
\bt{j}^{\chi}_\text{Hall}&=\frac{\sigma_\text{Hall}^{\chi}}{\text{E}_\text{0}\text{B}_\text{0}^2}(\bt{E} \cdot \bt{B}) \bt{B} \times \bt{E}.
\end{aligned}
\label{Acur2}
\end{equation}
If we set $\text{B}_\text{0}=1\text{ T}$ and $\text{E}_\text{0}=10^3 \text{ V/m}$, then
\begin{equation}
\begin{aligned}
\sigma_\text{D}^{\chi}&= \frac{1}{6}e^2 \frac{u\tau_\text{a}}{\pi^2 \hbar^3} (p_\text{F,pump}^{\chi})^2, \\
\frac{\sigma_\text{MR}^{\chi}}{[\text{T}^2]} &=\frac{1}{15} e^4 \frac{u\tau_\text{a}}{\pi^2 \hbar} \frac{1}{(p_\text{F,pump}^{\chi})^2},\\
\frac{\sigma_\text{Hall}^{\chi}}{[10^3 \text{ V/m} \cdot \text{T}^2]}&= \frac{1}{80} e^5 \frac{\tau_\text{a}}{\pi^2} \frac{\chi}{(p_\text{F,pump}^{\chi})^4}.
\end{aligned}
\label{Acurrents}
\end{equation}

Now, if we take a closer look, $ \bt{j}^{\chi}_\text{D}$ stems from the terms without Berry curvature,  
\begin{equation}
\begin{aligned}
e^2 \tau_a \pint  \bt{v}_\bt{p}^{\chi,\xi} (\bt{E}\cdot \nabla_{\bt{p}} f_0^{\chi,\xi})= \bt{j}^{\chi}_\text{D}.
\end{aligned}
\end{equation}
Containing the terms without cross product, gives $ \bt{j}^{\chi}_\text{D}$ and $ \bt{j}^{\chi}_\text{MR}$,
\begin{equation}
\begin{aligned}
e\tau_a \pint D^{\chi,\xi}(\bt{p})^{-1}  [ \bt{v}_\bt{p}^{\chi,\xi} + \frac{e}{\hbar} (\bp{\chi,\xi}\cdot \bt{v}_\bt{p}^{\chi,\xi})\bt{B} ] \\
[ e\bt{E}+ \frac{e^2}{\hbar} (\bt{E}\cdot \bt{B})\bp{\chi,\xi} ]\cdot \nabla_{\bt{p}} f_0^{\chi,\xi} = \bt{j}^{\chi}_\text{D} +  \bt{j}^{\chi}_\text{MR}.
\end{aligned}
\end{equation}
Finally, the term $ e \bt{v}_\bt{p}^{\chi,\xi} \times \bt{B} $ from $\dot{\bt{p}}^{\chi,\xi}$ does not bear any contribution, while $\frac{e}{\hbar} \bt{E} \times \bp{\chi,\xi}$ from $\dot{\bt{r}}^{\chi,\xi}$ results in $ \bt{j}^{\chi}_\text{Hall}$,
\begin{equation}
\begin{aligned}
\frac{e^2 \tau_a}{\hbar} \pint D^{\chi,\xi}(\bt{p})^{-1}  [\bt{E} \times \bp{\chi,\xi}]\\
[ e\bt{E}+ \frac{e^2}{\hbar} (\bt{E}\cdot \bt{B})\bp{\chi,\xi} ] \cdot \nabla_{\bt{p}} f_0^{\chi,\xi} = \bt{j}^{\chi}_\text{Hall}.
\end{aligned}
\end{equation}

\bibliographystyle{apsrev}

\begin{thebibliography}{0}
\expandafter\ifx\csname natexlab\endcsname\relax\def\natexlab#1{#1}\fi
\expandafter\ifx\csname bibnamefont\endcsname\relax
  \def\bibnamefont#1{#1}\fi
\expandafter\ifx\csname bibfnamefont\endcsname\relax
  \def\bibfnamefont#1{#1}\fi
\expandafter\ifx\csname citenamefont\endcsname\relax
  \def\citenamefont#1{#1}\fi
\expandafter\ifx\csname url\endcsname\relax
  \def\url#1{\texttt{#1}}\fi
\expandafter\ifx\csname urlprefix\endcsname\relax\def\urlprefix{URL }\fi
\providecommand{\bibinfo}[2]{#2}
\providecommand{\eprint}[2][]{\url{#2}}

\end{thebibliography}


\begin{thebibliography}{99}
\expandafter\ifx\csname natexlab\endcsname\relax\def\natexlab#1{#1}\fi
\expandafter\ifx\csname bibnamefont\endcsname\relax
  \def\bibnamefont#1{#1}\fi
\expandafter\ifx\csname bibfnamefont\endcsname\relax
  \def\bibfnamefont#1{#1}\fi
\expandafter\ifx\csname citenamefont\endcsname\relax
  \def\citenamefont#1{#1}\fi
\expandafter\ifx\csname url\endcsname\relax
  \def\url#1{\texttt{#1}}\fi
\expandafter\ifx\csname urlprefix\endcsname\relax\def\urlprefix{URL }\fi
\providecommand{\bibinfo}[2]{#2}
\providecommand{\eprint}[2][]{\url{#2}}

\bibitem[{\citenamefont{Xiao et~al.}(2010)\citenamefont{Xiao, Chang, and
  Niu}}]{Xiao1959}
\bibinfo{author}{\bibfnamefont{D.}~\bibnamefont{Xiao}},
  \bibinfo{author}{\bibfnamefont{M.-C.} \bibnamefont{Chang}}, \bibnamefont{and}
  \bibinfo{author}{\bibfnamefont{Q.}~\bibnamefont{Niu}}, \bibinfo{journal}{Rev.
  Mod. Phys.} \textbf{\bibinfo{volume}{82}}, \bibinfo{pages}{1959}
  (\bibinfo{year}{2010}).

\bibitem[{\citenamefont{Nagaosa et~al.}(2010)\citenamefont{Nagaosa, Sinova,
  Onoda, MacDonald, and Ong}}]{Nagaosa2010}
\bibinfo{author}{\bibfnamefont{N.}~\bibnamefont{Nagaosa}},
  \bibinfo{author}{\bibfnamefont{J.}~\bibnamefont{Sinova}},
  \bibinfo{author}{\bibfnamefont{S.}~\bibnamefont{Onoda}},
  \bibinfo{author}{\bibfnamefont{A.~H.} \bibnamefont{MacDonald}},
  \bibnamefont{and} \bibinfo{author}{\bibfnamefont{N.~P.} \bibnamefont{Ong}},
  \bibinfo{journal}{Rev. Mod. Phys.} \textbf{\bibinfo{volume}{82}},
  \bibinfo{pages}{1539} (\bibinfo{year}{2010}).

\bibitem[{\citenamefont{Qi and Zhang}(2011)}]{Qi2011}
\bibinfo{author}{\bibfnamefont{X.-L.} \bibnamefont{Qi}} \bibnamefont{and}
  \bibinfo{author}{\bibfnamefont{S.-C.} \bibnamefont{Zhang}},
  \bibinfo{journal}{Rev. Mod. Phys.} \textbf{\bibinfo{volume}{83}},
  \bibinfo{pages}{1057} (\bibinfo{year}{2011}).

\bibitem[{\citenamefont{Hosur and Qi}(2013)}]{Hosur2013}
\bibinfo{author}{\bibfnamefont{P.}~\bibnamefont{Hosur}} \bibnamefont{and}
  \bibinfo{author}{\bibfnamefont{X.}~\bibnamefont{Qi}},
  \bibinfo{journal}{Comptes Rendus Physique} \textbf{\bibinfo{volume}{14}},
  \bibinfo{pages}{857} (\bibinfo{year}{2013}).

\bibitem[{\citenamefont{Liang et~al.}(2015)\citenamefont{Liang, Gibson, Ali,
  Liu, Cava, and Ong}}]{Liang2015}
\bibinfo{author}{\bibfnamefont{T.}~\bibnamefont{Liang}},
  \bibinfo{author}{\bibfnamefont{Q.}~\bibnamefont{Gibson}},
  \bibinfo{author}{\bibfnamefont{M.~N.} \bibnamefont{Ali}},
  \bibinfo{author}{\bibfnamefont{M.}~\bibnamefont{Liu}},
  \bibinfo{author}{\bibfnamefont{R.~J.} \bibnamefont{Cava}}, \bibnamefont{and}
  \bibinfo{author}{\bibfnamefont{N.~P.} \bibnamefont{Ong}},
  \bibinfo{journal}{Nature materials} \textbf{\bibinfo{volume}{14}},
  \bibinfo{pages}{280} (\bibinfo{year}{2015}).

\bibitem[{\citenamefont{Liang et~al.}(2018)\citenamefont{Liang, Lin, Gibson,
  Kushwaha, Liu, Wang, Xiong, Sobota, Hashimoto, Kirchmann et~al.}}]{Liang2018}
\bibinfo{author}{\bibfnamefont{T.}~\bibnamefont{Liang}},
  \bibinfo{author}{\bibfnamefont{J.}~\bibnamefont{Lin}},
  \bibinfo{author}{\bibfnamefont{Q.}~\bibnamefont{Gibson}},
  \bibinfo{author}{\bibfnamefont{S.}~\bibnamefont{Kushwaha}},
  \bibinfo{author}{\bibfnamefont{M.}~\bibnamefont{Liu}},
  \bibinfo{author}{\bibfnamefont{W.}~\bibnamefont{Wang}},
  \bibinfo{author}{\bibfnamefont{H.}~\bibnamefont{Xiong}},
  \bibinfo{author}{\bibfnamefont{J.~A.} \bibnamefont{Sobota}},
  \bibinfo{author}{\bibfnamefont{M.}~\bibnamefont{Hashimoto}},
  \bibinfo{author}{\bibfnamefont{P.~S.} \bibnamefont{Kirchmann}},
  \bibnamefont{et~al.}, \bibinfo{journal}{Nature Physics}
  \textbf{\bibinfo{volume}{14}}, \bibinfo{pages}{451} (\bibinfo{year}{2018}).

\bibitem[{\citenamefont{Armitage et~al.}(2018)\citenamefont{Armitage, Mele, and
  Vishwanath}}]{Armitage2018}
\bibinfo{author}{\bibfnamefont{N.~P.} \bibnamefont{Armitage}},
  \bibinfo{author}{\bibfnamefont{E.~J.} \bibnamefont{Mele}}, \bibnamefont{and}
  \bibinfo{author}{\bibfnamefont{A.}~\bibnamefont{Vishwanath}},
  \bibinfo{journal}{Reviews of Modern Physics} \textbf{\bibinfo{volume}{90}},
  \bibinfo{pages}{015001} (\bibinfo{year}{2018}).

\bibitem[{\citenamefont{Moore and Balents}(2007)}]{Moore2007}
\bibinfo{author}{\bibfnamefont{J.~E.} \bibnamefont{Moore}} \bibnamefont{and}
  \bibinfo{author}{\bibfnamefont{L.}~\bibnamefont{Balents}},
  \bibinfo{journal}{Physical Review B} \textbf{\bibinfo{volume}{75}},
  \bibinfo{pages}{121306(R)} (\bibinfo{year}{2007}).

\bibitem[{\citenamefont{Fu et~al.}(2007)\citenamefont{Fu, Kane, and
  Mele}}]{Fu2007}
\bibinfo{author}{\bibfnamefont{L.}~\bibnamefont{Fu}},
  \bibinfo{author}{\bibfnamefont{C.~L.} \bibnamefont{Kane}}, \bibnamefont{and}
  \bibinfo{author}{\bibfnamefont{E.~J.} \bibnamefont{Mele}},
  \bibinfo{journal}{Physical review letters} \textbf{\bibinfo{volume}{98}},
  \bibinfo{pages}{106803} (\bibinfo{year}{2007}).

\bibitem[{\citenamefont{Hasan and Kane}(2010)}]{Hasan2010}
\bibinfo{author}{\bibfnamefont{M.~Z.} \bibnamefont{Hasan}} \bibnamefont{and}
  \bibinfo{author}{\bibfnamefont{C.~L.} \bibnamefont{Kane}},
  \bibinfo{journal}{Rev. Mod. Phys.} \textbf{\bibinfo{volume}{82}},
  \bibinfo{pages}{3045} (\bibinfo{year}{2010}).

\bibitem[{\citenamefont{Weyl}(1929)}]{Weyl1929}
\bibinfo{author}{\bibfnamefont{H.}~\bibnamefont{Weyl}}, \bibinfo{journal}{z.
  Phys.} \textbf{\bibinfo{volume}{56}}, \bibinfo{pages}{330}
  (\bibinfo{year}{1929}).

\bibitem[{\citenamefont{Herring}(1937)}]{Herring1937}
\bibinfo{author}{\bibfnamefont{C.}~\bibnamefont{Herring}},
  \bibinfo{journal}{Physical Review} \textbf{\bibinfo{volume}{52}},
  \bibinfo{pages}{365} (\bibinfo{year}{1937}).

\bibitem[{\citenamefont{Murakami}(2007)}]{Murakami2007}
\bibinfo{author}{\bibfnamefont{S.}~\bibnamefont{Murakami}},
  \bibinfo{journal}{New Journal of Physics} \textbf{\bibinfo{volume}{9}},
  \bibinfo{pages}{356} (\bibinfo{year}{2007}).

\bibitem[{\citenamefont{Burkov and Balents}(2011)}]{Burkov2011}
\bibinfo{author}{\bibfnamefont{A.~A.} \bibnamefont{Burkov}} \bibnamefont{and}
  \bibinfo{author}{\bibfnamefont{L.}~\bibnamefont{Balents}},
  \bibinfo{journal}{Physical review letters} \textbf{\bibinfo{volume}{107}},
  \bibinfo{pages}{127205} (\bibinfo{year}{2011}).

\bibitem[{\citenamefont{Huang et~al.}(2015{\natexlab{a}})\citenamefont{Huang,
  Xu, Belopolski, Lee, Chang, Wang, Alidoust, Bian, Neupane, Zhang
  et~al.}}]{Huang2015}
\bibinfo{author}{\bibfnamefont{S.-M.} \bibnamefont{Huang}},
  \bibinfo{author}{\bibfnamefont{S.-Y.} \bibnamefont{Xu}},
  \bibinfo{author}{\bibfnamefont{I.}~\bibnamefont{Belopolski}},
  \bibinfo{author}{\bibfnamefont{C.-C.} \bibnamefont{Lee}},
  \bibinfo{author}{\bibfnamefont{G.}~\bibnamefont{Chang}},
  \bibinfo{author}{\bibfnamefont{B.}~\bibnamefont{Wang}},
  \bibinfo{author}{\bibfnamefont{N.}~\bibnamefont{Alidoust}},
  \bibinfo{author}{\bibfnamefont{G.}~\bibnamefont{Bian}},
  \bibinfo{author}{\bibfnamefont{M.}~\bibnamefont{Neupane}},
  \bibinfo{author}{\bibfnamefont{C.}~\bibnamefont{Zhang}},
  \bibnamefont{et~al.}, \bibinfo{journal}{Nature communications}
  \textbf{\bibinfo{volume}{6}}, \bibinfo{pages}{7373}
  (\bibinfo{year}{2015}{\natexlab{a}}).

\bibitem[{\citenamefont{Burkov}(2018)}]{Burkov2018}
\bibinfo{author}{\bibfnamefont{A.~A.} \bibnamefont{Burkov}},
  \bibinfo{journal}{Annual Review of Condensed Matter Physics}
  \textbf{\bibinfo{volume}{9}}, \bibinfo{pages}{359} (\bibinfo{year}{2018}).

\bibitem[{\citenamefont{Adler}(1969)}]{Adler1969}
\bibinfo{author}{\bibfnamefont{S.~L.} \bibnamefont{Adler}},
  \bibinfo{journal}{Physical Review} \textbf{\bibinfo{volume}{177}},
  \bibinfo{pages}{2426} (\bibinfo{year}{1969}).

\bibitem[{\citenamefont{Bell and Jackiw}(1969)}]{Bell1969}
\bibinfo{author}{\bibfnamefont{J.~S.} \bibnamefont{Bell}} \bibnamefont{and}
  \bibinfo{author}{\bibfnamefont{R.}~\bibnamefont{Jackiw}},
  \bibinfo{journal}{Il Nuovo Cimento A (1965-1970)}
  \textbf{\bibinfo{volume}{60}}, \bibinfo{pages}{47} (\bibinfo{year}{1969}).

\bibitem[{\citenamefont{Aji}(2012)}]{Aji2012}
\bibinfo{author}{\bibfnamefont{V.}~\bibnamefont{Aji}},
  \bibinfo{journal}{Physical Review B} \textbf{\bibinfo{volume}{85}},
  \bibinfo{pages}{241101(R)} (\bibinfo{year}{2012}).

\bibitem[{\citenamefont{Son and Spivak}(2013)}]{Son2013}
\bibinfo{author}{\bibfnamefont{D.~T.} \bibnamefont{Son}} \bibnamefont{and}
  \bibinfo{author}{\bibfnamefont{B.~Z.} \bibnamefont{Spivak}},
  \bibinfo{journal}{Phys. Rev. B} \textbf{\bibinfo{volume}{88}},
  \bibinfo{pages}{104412} (\bibinfo{year}{2013}).

\bibitem[{\citenamefont{Fukushima et~al.}(2008)\citenamefont{Fukushima,
  Kharzeev, and Warringa}}]{Fukushima2008}
\bibinfo{author}{\bibfnamefont{K.}~\bibnamefont{Fukushima}},
  \bibinfo{author}{\bibfnamefont{D.~E.} \bibnamefont{Kharzeev}},
  \bibnamefont{and} \bibinfo{author}{\bibfnamefont{H.~J.}
  \bibnamefont{Warringa}}, \bibinfo{journal}{Physical Review D}
  \textbf{\bibinfo{volume}{78}}, \bibinfo{pages}{074033}
  (\bibinfo{year}{2008}).

\bibitem[{\citenamefont{Kharzeev and Warringa}(2009)}]{Kharzeev2009}
\bibinfo{author}{\bibfnamefont{D.~E.} \bibnamefont{Kharzeev}} \bibnamefont{and}
  \bibinfo{author}{\bibfnamefont{H.~J.} \bibnamefont{Warringa}},
  \bibinfo{journal}{Physical Review D} \textbf{\bibinfo{volume}{80}},
  \bibinfo{pages}{034028} (\bibinfo{year}{2009}).

\bibitem[{\citenamefont{Son and Yamamoto}(2012)}]{Son2012}
\bibinfo{author}{\bibfnamefont{D.~T.} \bibnamefont{Son}} \bibnamefont{and}
  \bibinfo{author}{\bibfnamefont{N.}~\bibnamefont{Yamamoto}},
  \bibinfo{journal}{Physical review letters} \textbf{\bibinfo{volume}{109}},
  \bibinfo{pages}{181602} (\bibinfo{year}{2012}).

\bibitem[{\citenamefont{Stephanov and Yin}(2012)}]{Stephanov2012}
\bibinfo{author}{\bibfnamefont{M.~A.} \bibnamefont{Stephanov}}
  \bibnamefont{and} \bibinfo{author}{\bibfnamefont{Y.}~\bibnamefont{Yin}},
  \bibinfo{journal}{Physical review letters} \textbf{\bibinfo{volume}{109}},
  \bibinfo{pages}{162001} (\bibinfo{year}{2012}).

\bibitem[{\citenamefont{Burkov}(2015{\natexlab{a}})}]{Burkov2015prb}
\bibinfo{author}{\bibfnamefont{A.~A.} \bibnamefont{Burkov}},
  \bibinfo{journal}{Physical Review B} \textbf{\bibinfo{volume}{91}},
  \bibinfo{pages}{245157} (\bibinfo{year}{2015}{\natexlab{a}}).

\bibitem[{\citenamefont{Li et~al.}(2016)\citenamefont{Li, Kharzeev, Zhang,
  Huang, Pletikosi{\'c}, Fedorov, Zhong, Schneeloch, Gu, and Valla}}]{Li2016}
\bibinfo{author}{\bibfnamefont{Q.}~\bibnamefont{Li}},
  \bibinfo{author}{\bibfnamefont{D.~E.} \bibnamefont{Kharzeev}},
  \bibinfo{author}{\bibfnamefont{C.}~\bibnamefont{Zhang}},
  \bibinfo{author}{\bibfnamefont{Y.}~\bibnamefont{Huang}},
  \bibinfo{author}{\bibfnamefont{I.}~\bibnamefont{Pletikosi{\'c}}},
  \bibinfo{author}{\bibfnamefont{A.~V.} \bibnamefont{Fedorov}},
  \bibinfo{author}{\bibfnamefont{R.~D.} \bibnamefont{Zhong}},
  \bibinfo{author}{\bibfnamefont{J.~A.} \bibnamefont{Schneeloch}},
  \bibinfo{author}{\bibfnamefont{G.~D.} \bibnamefont{Gu}}, \bibnamefont{and}
  \bibinfo{author}{\bibfnamefont{T.}~\bibnamefont{Valla}},
  \bibinfo{journal}{Nature Physics} \textbf{\bibinfo{volume}{12}},
  \bibinfo{pages}{550} (\bibinfo{year}{2016}).
    
\bibitem[{\citenamefont{Cheon et~al.}(2022)\citenamefont{Cheon, Cho, Kim, and
  Lee}}]{Cheon2022}
\bibinfo{author}{\bibfnamefont{S.}~\bibnamefont{Cheon}},
  \bibinfo{author}{\bibfnamefont{G.~Y.} \bibnamefont{Cho}},
  \bibinfo{author}{\bibfnamefont{K.-S.} \bibnamefont{Kim}}, \bibnamefont{and}
  \bibinfo{author}{\bibfnamefont{H.-W.} \bibnamefont{Lee}},
  \bibinfo{journal}{Phys. Rev. B} \textbf{\bibinfo{volume}{105}},
  \bibinfo{pages}{L180303} (\bibinfo{year}{2022}).

\bibitem[{\citenamefont{Nielsen and Ninomiya}(1983)}]{Nielsen1983}
\bibinfo{author}{\bibfnamefont{H.~B.} \bibnamefont{Nielsen}} \bibnamefont{and}
  \bibinfo{author}{\bibfnamefont{M.}~\bibnamefont{Ninomiya}},
  \bibinfo{journal}{Physics Letters B} \textbf{\bibinfo{volume}{130}},
  \bibinfo{pages}{389} (\bibinfo{year}{1983}).

\bibitem[{\citenamefont{Kim et~al.}(2013)\citenamefont{Kim, Kim, Wang, Sasaki,
  Satoh, Ohnishi, Kitaura, Yang, and Li}}]{Kim2013}
\bibinfo{author}{\bibfnamefont{H.-J.} \bibnamefont{Kim}},
  \bibinfo{author}{\bibfnamefont{K.-S.} \bibnamefont{Kim}},
  \bibinfo{author}{\bibfnamefont{J.~F.} \bibnamefont{Wang}},
  \bibinfo{author}{\bibfnamefont{M.}~\bibnamefont{Sasaki}},
  \bibinfo{author}{\bibfnamefont{N.}~\bibnamefont{Satoh}},
  \bibinfo{author}{\bibfnamefont{A.}~\bibnamefont{Ohnishi}},
  \bibinfo{author}{\bibfnamefont{M.}~\bibnamefont{Kitaura}},
  \bibinfo{author}{\bibfnamefont{M.}~\bibnamefont{Yang}}, \bibnamefont{and}
  \bibinfo{author}{\bibfnamefont{L.}~\bibnamefont{Li}},
  \bibinfo{journal}{Physical review letters} \textbf{\bibinfo{volume}{111}},
  \bibinfo{pages}{246603} (\bibinfo{year}{2013}).

\bibitem[{\citenamefont{Xiong et~al.}(2015)\citenamefont{Xiong, Kushwaha,
  Liang, Krizan, Hirschberger, Wang, Cava, and Ong}}]{Xiong2015}
\bibinfo{author}{\bibfnamefont{J.}~\bibnamefont{Xiong}},
  \bibinfo{author}{\bibfnamefont{S.~K.} \bibnamefont{Kushwaha}},
  \bibinfo{author}{\bibfnamefont{T.}~\bibnamefont{Liang}},
  \bibinfo{author}{\bibfnamefont{J.~W.} \bibnamefont{Krizan}},
  \bibinfo{author}{\bibfnamefont{M.}~\bibnamefont{Hirschberger}},
  \bibinfo{author}{\bibfnamefont{W.}~\bibnamefont{Wang}},
  \bibinfo{author}{\bibfnamefont{R.~J.} \bibnamefont{Cava}}, \bibnamefont{and}
  \bibinfo{author}{\bibfnamefont{N.~P.} \bibnamefont{Ong}},
  \bibinfo{journal}{Science} \textbf{\bibinfo{volume}{350}},
  \bibinfo{pages}{413} (\bibinfo{year}{2015}).

\bibitem[{\citenamefont{Huang et~al.}(2015{\natexlab{b}})\citenamefont{Huang,
  Zhao, Long, Wang, Chen, Yang, Liang, Xue, Weng, Fang et~al.}}]{HuangX2015}
\bibinfo{author}{\bibfnamefont{X.}~\bibnamefont{Huang}},
  \bibinfo{author}{\bibfnamefont{L.}~\bibnamefont{Zhao}},
  \bibinfo{author}{\bibfnamefont{Y.}~\bibnamefont{Long}},
  \bibinfo{author}{\bibfnamefont{P.}~\bibnamefont{Wang}},
  \bibinfo{author}{\bibfnamefont{D.}~\bibnamefont{Chen}},
  \bibinfo{author}{\bibfnamefont{Z.}~\bibnamefont{Yang}},
  \bibinfo{author}{\bibfnamefont{H.}~\bibnamefont{Liang}},
  \bibinfo{author}{\bibfnamefont{M.}~\bibnamefont{Xue}},
  \bibinfo{author}{\bibfnamefont{H.}~\bibnamefont{Weng}},
  \bibinfo{author}{\bibfnamefont{Z.}~\bibnamefont{Fang}},
  \bibinfo{author}{\bibfnamefont{X.}~\bibnamefont{Dai}},
  \bibinfo{author}{\bibfnamefont{G.}~\bibnamefont{Chen}}, 
  \bibinfo{journal}{Phys. Rev. X} \textbf{\bibinfo{volume}{5}},
  \bibinfo{pages}{031023} (\bibinfo{year}{2015}{\natexlab{b}}).

\bibitem[{\citenamefont{Kim et~al.}(2009)\citenamefont{Kim, Choi, and
  Min}}]{Kim2009}
\bibinfo{author}{\bibfnamefont{K.}~\bibnamefont{Kim}},
  \bibinfo{author}{\bibfnamefont{H.~C.} \bibnamefont{Choi}}, \bibnamefont{and}
  \bibinfo{author}{\bibfnamefont{B.~I.} \bibnamefont{Min}},
  \bibinfo{journal}{Physical Review B} \textbf{\bibinfo{volume}{80}},
  \bibinfo{pages}{035116} (\bibinfo{year}{2009}).

\bibitem[{\citenamefont{Noh et~al.}(2009)\citenamefont{Noh, Jeong, Jeong, Cho,
  Kim, Kim, Min, and Kim}}]{Noh2009}
\bibinfo{author}{\bibfnamefont{H.-J.} \bibnamefont{Noh}},
  \bibinfo{author}{\bibfnamefont{J.}~\bibnamefont{Jeong}},
  \bibinfo{author}{\bibfnamefont{J.}~\bibnamefont{Jeong}},
  \bibinfo{author}{\bibfnamefont{E.-J.} \bibnamefont{Cho}},
  \bibinfo{author}{\bibfnamefont{S.~B.} \bibnamefont{Kim}},
  \bibinfo{author}{\bibfnamefont{K.}~\bibnamefont{Kim}},
  \bibinfo{author}{\bibfnamefont{B.~I.} \bibnamefont{Min}}, \bibnamefont{and}
  \bibinfo{author}{\bibfnamefont{H.-D.} \bibnamefont{Kim}},
  \bibinfo{journal}{Physical review letters} \textbf{\bibinfo{volume}{102}},
  \bibinfo{pages}{256404} (\bibinfo{year}{2009}).

\bibitem[{\citenamefont{Goswami et~al.}(2015)\citenamefont{Goswami, Pixley, and
  Das~Sarma}}]{Goswami2015}
\bibinfo{author}{\bibfnamefont{P.}~\bibnamefont{Goswami}},
  \bibinfo{author}{\bibfnamefont{J.~H.} \bibnamefont{Pixley}},
  \bibnamefont{and}
  \bibinfo{author}{\bibfnamefont{S.}~\bibnamefont{Das~Sarma}},
  \bibinfo{journal}{Physical Review B} \textbf{\bibinfo{volume}{92}},
  \bibinfo{pages}{075205} (\bibinfo{year}{2015}).

\bibitem[{\citenamefont{Dos~Reis et~al.}(2016)\citenamefont{Dos~Reis, Ajeesh,
  Kumar, Arnold, Shekhar, Naumann, Schmidt, Nicklas, and Hassinger}}]{Dos2016}
\bibinfo{author}{\bibfnamefont{R.~D.} \bibnamefont{Dos~Reis}},
  \bibinfo{author}{\bibfnamefont{M.~O.} \bibnamefont{Ajeesh}},
  \bibinfo{author}{\bibfnamefont{N.}~\bibnamefont{Kumar}},
  \bibinfo{author}{\bibfnamefont{F.}~\bibnamefont{Arnold}},
  \bibinfo{author}{\bibfnamefont{C.}~\bibnamefont{Shekhar}},
  \bibinfo{author}{\bibfnamefont{M.}~\bibnamefont{Naumann}},
  \bibinfo{author}{\bibfnamefont{M.}~\bibnamefont{Schmidt}},
  \bibinfo{author}{\bibfnamefont{M.}~\bibnamefont{Nicklas}}, \bibnamefont{and}
  \bibinfo{author}{\bibfnamefont{E.}~\bibnamefont{Hassinger}},
  \bibinfo{journal}{New Journal of Physics} \textbf{\bibinfo{volume}{18}},
  \bibinfo{pages}{085006} (\bibinfo{year}{2016}).

\bibitem[{\citenamefont{Kikugawa et~al.}(2016)\citenamefont{Kikugawa, Goswami,
  Kiswandhi, Choi, Graf, Baumbach, Brooks, Sugii, Iida, Nishio
  et~al.}}]{Kikugawa2016}
\bibinfo{author}{\bibfnamefont{N.}~\bibnamefont{Kikugawa}},
  \bibinfo{author}{\bibfnamefont{P.}~\bibnamefont{Goswami}},
  \bibinfo{author}{\bibfnamefont{A.}~\bibnamefont{Kiswandhi}},
  \bibinfo{author}{\bibfnamefont{E.~S.} \bibnamefont{Choi}},
  \bibinfo{author}{\bibfnamefont{D.}~\bibnamefont{Graf}},
  \bibinfo{author}{\bibfnamefont{R.~E.} \bibnamefont{Baumbach}},
  \bibinfo{author}{\bibfnamefont{J.~S.} \bibnamefont{Brooks}},
  \bibinfo{author}{\bibfnamefont{K.}~\bibnamefont{Sugii}},
  \bibinfo{author}{\bibfnamefont{Y.}~\bibnamefont{Iida}},
  \bibinfo{author}{\bibfnamefont{M.}~\bibnamefont{Nishio}},
  \bibnamefont{et~al.}, \bibinfo{journal}{Nature communications}
  \textbf{\bibinfo{volume}{7}}, \bibinfo{pages}{10903} (\bibinfo{year}{2016}).

\bibitem[{\citenamefont{Ong and Liang}(2021)}]{Ong2021}
\bibinfo{author}{\bibfnamefont{N.~P.} \bibnamefont{Ong}} \bibnamefont{and}
  \bibinfo{author}{\bibfnamefont{S.}~\bibnamefont{Liang}},
  \bibinfo{journal}{Nature Reviews Physics} \textbf{\bibinfo{volume}{3}},
  \bibinfo{pages}{394} (\bibinfo{year}{2021}).

\bibitem[{\citenamefont{Morimoto and Nagaosa}(2016)}]{Morimoto2016}
\bibinfo{author}{\bibfnamefont{T.}~\bibnamefont{Morimoto}} \bibnamefont{and}
  \bibinfo{author}{\bibfnamefont{N.}~\bibnamefont{Nagaosa}},
  \bibinfo{journal}{Physical Review Letters} \textbf{\bibinfo{volume}{117}},
  \bibinfo{pages}{146603} (\bibinfo{year}{2016}).

\bibitem[{\citenamefont{Shin et~al.}(2017)\citenamefont{Shin, Lee, Sasaki,
  Jeong, Weickert, Betts, Kim, Kim, and Kim}}]{Shin2017}
\bibinfo{author}{\bibfnamefont{D.}~\bibnamefont{Shin}},
  \bibinfo{author}{\bibfnamefont{Y.}~\bibnamefont{Lee}},
  \bibinfo{author}{\bibfnamefont{M.}~\bibnamefont{Sasaki}},
  \bibinfo{author}{\bibfnamefont{Y.~H.} \bibnamefont{Jeong}},
  \bibinfo{author}{\bibfnamefont{F.}~\bibnamefont{Weickert}},
  \bibinfo{author}{\bibfnamefont{J.~B.} \bibnamefont{Betts}},
  \bibinfo{author}{\bibfnamefont{H.-J.} \bibnamefont{Kim}},
  \bibinfo{author}{\bibfnamefont{K.-S.} \bibnamefont{Kim}}, \bibnamefont{and}
  \bibinfo{author}{\bibfnamefont{J.}~\bibnamefont{Kim}},
  \bibinfo{journal}{Nature materials} \textbf{\bibinfo{volume}{16}},
  \bibinfo{pages}{1096} (\bibinfo{year}{2017}).

\bibitem[{\citenamefont{Nagpal and Patnaik}(2020)}]{Nagpal2020}
\bibinfo{author}{\bibfnamefont{V.}~\bibnamefont{Nagpal}} \bibnamefont{and}
  \bibinfo{author}{\bibfnamefont{S.}~\bibnamefont{Patnaik}},
  \bibinfo{journal}{Journal of Physics: Condensed Matter}
  \textbf{\bibinfo{volume}{32}}, \bibinfo{pages}{405602}
  (\bibinfo{year}{2020}).

\bibitem[{\citenamefont{Nandy et~al.}(2021)\citenamefont{Nandy, Zeng, and
  Tewari}}]{Nandy2021}
\bibinfo{author}{\bibfnamefont{S.}~\bibnamefont{Nandy}},
  \bibinfo{author}{\bibfnamefont{C.}~\bibnamefont{Zeng}}, \bibnamefont{and}
  \bibinfo{author}{\bibfnamefont{S.}~\bibnamefont{Tewari}},
  \bibinfo{journal}{Physical Review B} \textbf{\bibinfo{volume}{104}},
  \bibinfo{pages}{205124} (\bibinfo{year}{2021}).

\bibitem[{\citenamefont{Li et~al.}(2021)\citenamefont{Li, Heinonen, Burkov, and
  Zhang}}]{Li2021}
\bibinfo{author}{\bibfnamefont{R.-H.} \bibnamefont{Li}},
  \bibinfo{author}{\bibfnamefont{O.~G.} \bibnamefont{Heinonen}},
  \bibinfo{author}{\bibfnamefont{A.~A.} \bibnamefont{Burkov}},
  \bibnamefont{and} \bibinfo{author}{\bibfnamefont{S.~S.~-L.}
  \bibnamefont{Zhang}}, \bibinfo{journal}{Physical Review B}
  \textbf{\bibinfo{volume}{103}}, \bibinfo{pages}{045105}
  (\bibinfo{year}{2021}).

\bibitem[{\citenamefont{Vashist et~al.}(2021)\citenamefont{Vashist, Gopal, and
  Singh}}]{Vashist2021}
\bibinfo{author}{\bibfnamefont{A.}~\bibnamefont{Vashist}},
  \bibinfo{author}{\bibfnamefont{R.~K.} \bibnamefont{Gopal}}, \bibnamefont{and}
  \bibinfo{author}{\bibfnamefont{Y.}~\bibnamefont{Singh}},
  \bibinfo{journal}{Scientific reports} \textbf{\bibinfo{volume}{11}},
  \bibinfo{pages}{1} (\bibinfo{year}{2021}).

\bibitem[{\citenamefont{Kim et~al.}(2014)\citenamefont{Kim, Kim, and
  Sasaki}}]{Kim2014}
\bibinfo{author}{\bibfnamefont{K.-S.} \bibnamefont{Kim}},
  \bibinfo{author}{\bibfnamefont{H.-J.} \bibnamefont{Kim}}, \bibnamefont{and}
  \bibinfo{author}{\bibfnamefont{M.}~\bibnamefont{Sasaki}},
  \bibinfo{journal}{Physical Review B} \textbf{\bibinfo{volume}{89}},
  \bibinfo{pages}{195137} (\bibinfo{year}{2014}).

\bibitem[{\citenamefont{Weng et~al.}(2015)\citenamefont{Weng, Fang, Fang,
  Bernevig, and Dai}}]{Weng2015}
\bibinfo{author}{\bibfnamefont{H.}~\bibnamefont{Weng}},
  \bibinfo{author}{\bibfnamefont{C.}~\bibnamefont{Fang}},
  \bibinfo{author}{\bibfnamefont{Z.}~\bibnamefont{Fang}},
  \bibinfo{author}{\bibfnamefont{B.~A.} \bibnamefont{Bernevig}},
  \bibnamefont{and} \bibinfo{author}{\bibfnamefont{X.}~\bibnamefont{Dai}},
  \bibinfo{journal}{Physical Review X} \textbf{\bibinfo{volume}{5}},
  \bibinfo{pages}{011029} (\bibinfo{year}{2015}).

\bibitem[{\citenamefont{Ramshaw et~al.}(2018)\citenamefont{Ramshaw, Modic,
  Shekhter, Zhang, Kim, Moll, Bachmann, Chan, Betts, Balakirev
  et~al.}}]{Ramshaw2018}
\bibinfo{author}{\bibfnamefont{B.~J.} \bibnamefont{Ramshaw}},
  \bibinfo{author}{\bibfnamefont{K.~A.} \bibnamefont{Modic}},
  \bibinfo{author}{\bibfnamefont{A.}~\bibnamefont{Shekhter}},
  \bibinfo{author}{\bibfnamefont{Y.}~\bibnamefont{Zhang}},
  \bibinfo{author}{\bibfnamefont{E.-A.} \bibnamefont{Kim}},
  \bibinfo{author}{\bibfnamefont{P.~J.~W.} \bibnamefont{Moll}},
  \bibinfo{author}{\bibfnamefont{M.~D.} \bibnamefont{Bachmann}},
  \bibinfo{author}{\bibfnamefont{M.~K.} \bibnamefont{Chan}},
  \bibinfo{author}{\bibfnamefont{J.~B.} \bibnamefont{Betts}},
  \bibinfo{author}{\bibfnamefont{F.}~\bibnamefont{Balakirev}},
  \bibnamefont{et~al.}, \bibinfo{journal}{Nature communications}
  \textbf{\bibinfo{volume}{9}}, \bibinfo{pages}{2217} (\bibinfo{year}{2018}).

\bibitem[{\citenamefont{Burkov}(2015{\natexlab{b}})}]{Burkov2015iop}
\bibinfo{author}{\bibfnamefont{A.~A.} \bibnamefont{Burkov}},
  \bibinfo{journal}{Journal of Physics: Condensed Matter}
  \textbf{\bibinfo{volume}{27}}, \bibinfo{pages}{113201}
  (\bibinfo{year}{2015}{\natexlab{b}}).

\bibitem[{\citenamefont{Vazifeh and Franz}(2013)}]{Vazifeh2013}
\bibinfo{author}{\bibfnamefont{M.~M.} \bibnamefont{Vazifeh}} \bibnamefont{and}
  \bibinfo{author}{\bibfnamefont{M.}~\bibnamefont{Franz}},
  \bibinfo{journal}{Physical review letters} \textbf{\bibinfo{volume}{111}},
  \bibinfo{pages}{027201} (\bibinfo{year}{2013}).

\bibitem[{\citenamefont{Goswami and Tewari}(2013)}]{Goswami2013}
\bibinfo{author}{\bibfnamefont{P.}~\bibnamefont{Goswami}} \bibnamefont{and}
  \bibinfo{author}{\bibfnamefont{S.}~\bibnamefont{Tewari}},
  \bibinfo{journal}{arXiv preprint arXiv:1311.1506}  (\bibinfo{year}{2013}).

\bibitem[{\citenamefont{Xiao et~al.}(2005)\citenamefont{Xiao, Shi, and
  Niu}}]{Xiao2005}
\bibinfo{author}{\bibfnamefont{D.}~\bibnamefont{Xiao}},
  \bibinfo{author}{\bibfnamefont{J.}~\bibnamefont{Shi}}, \bibnamefont{and}
  \bibinfo{author}{\bibfnamefont{Q.}~\bibnamefont{Niu}},
  \bibinfo{journal}{Physical review letters} \textbf{\bibinfo{volume}{95}},
  \bibinfo{pages}{137204} (\bibinfo{year}{2005}).

\bibitem[{\citenamefont{Sekine and Nomura}(2021)}]{Sekine2021}
\bibinfo{author}{\bibfnamefont{A.}~\bibnamefont{Sekine}} \bibnamefont{and}
  \bibinfo{author}{\bibfnamefont{K.}~\bibnamefont{Nomura}},
  \bibinfo{journal}{Journal of Applied Physics} \textbf{\bibinfo{volume}{129}},
  \bibinfo{pages}{141101} (\bibinfo{year}{2021}).

\bibitem[{\citenamefont{Takasan et~al.}(2021)\citenamefont{Takasan, Morimoto,
  Orenstein, and Moore}}]{Takasan2021}
\bibinfo{author}{\bibfnamefont{K.}~\bibnamefont{Takasan}},
  \bibinfo{author}{\bibfnamefont{T.}~\bibnamefont{Morimoto}},
  \bibinfo{author}{\bibfnamefont{J.}~\bibnamefont{Orenstein}},
  \bibnamefont{and} \bibinfo{author}{\bibfnamefont{J.~E.} \bibnamefont{Moore}},
  \bibinfo{journal}{Physical Review B} \textbf{\bibinfo{volume}{104}},
  \bibinfo{pages}{L161202} (\bibinfo{year}{2021}).

\bibitem[{\citenamefont{Sundaram and Niu}(1999)}]{Sundaram1999}
\bibinfo{author}{\bibfnamefont{G.}~\bibnamefont{Sundaram}} \bibnamefont{and}
  \bibinfo{author}{\bibfnamefont{Q.}~\bibnamefont{Niu}},
  \bibinfo{journal}{Physical Review B} \textbf{\bibinfo{volume}{59}},
  \bibinfo{pages}{14915} (\bibinfo{year}{1999}).

\bibitem[{\citenamefont{Zhang et~al.}(2016)\citenamefont{Zhang, Xu, Belopolski,
  Yuan, Lin, Tong, Bian, Alidoust, Lee, Huang et~al.}}]{Zhang2016}
\bibinfo{author}{\bibfnamefont{C.-L.} \bibnamefont{Zhang}},
  \bibinfo{author}{\bibfnamefont{S.-Y.} \bibnamefont{Xu}},
  \bibinfo{author}{\bibfnamefont{I.}~\bibnamefont{Belopolski}},
  \bibinfo{author}{\bibfnamefont{Z.}~\bibnamefont{Yuan}},
  \bibinfo{author}{\bibfnamefont{Z.}~\bibnamefont{Lin}},
  \bibinfo{author}{\bibfnamefont{B.}~\bibnamefont{Tong}},
  \bibinfo{author}{\bibfnamefont{G.}~\bibnamefont{Bian}},
  \bibinfo{author}{\bibfnamefont{N.}~\bibnamefont{Alidoust}},
  \bibinfo{author}{\bibfnamefont{C.-C.} \bibnamefont{Lee}},
  \bibinfo{author}{\bibfnamefont{S.-M.} \bibnamefont{Huang}},
  \bibnamefont{et~al.}, \bibinfo{journal}{Nature communications}
  \textbf{\bibinfo{volume}{7}}, \bibinfo{pages}{10735} (\bibinfo{year}{2016}).



\end{thebibliography}

\end{document}